\documentclass[fleqn,usenatbib]{mnras}

\usepackage[T1]{fontenc}

\DeclareRobustCommand{\VAN}[3]{#2}
\let\VANthebibliography\thebibliography
\def\thebibliography{\DeclareRobustCommand{\VAN}[3]{##3}\VANthebibliography}

\usepackage{graphicx}	
\usepackage{amsmath}	
\usepackage{amssymb}	

\usepackage{newtxtext,newtxmath}

\usepackage{tikz}

\newcommand{\avg}[1]{\left\langle{#1}\right\rangle}

\newif\iftrack
\iftrack
\newcommand{\added}[1]{{\bf #1}}
\newcommand{\deleted}[1]{}
\newcommand{\replaced}[2]{{\bf #2}}
\else
\newcommand{\added}[1]{{#1}}
\newcommand{\deleted}[1]{}
\newcommand{\replaced}[2]{{#2}}
\fi

\title[COMAP-ERA DDE forecasts]{The deconvolved distribution estimator: enhancing reionisation-era CO line-intensity mapping analyses with a cross-correlation analogue for one-point statistics}

\author[D.~T.~Chung et al.]{Dongwoo T.~Chung,$^{1,2}$\thanks{E-mail: dongwooc@cita.utoronto.ca} Ishika Bangari,$^3$ Patrick C.~Breysse,$^4$ H\aa vard T.~Ihle,$^5$ J.~Richard Bond,$^{1}$ \newauthor Delaney A.~Dunne,$^{6}$ Hamsa Padmanabhan,$^{7}$ Liju Philip,$^{6}$ Thomas J.~Rennie,$^{8}$ and Marco P.~Viero$^{6}$\newauthor (COMAP Collaboration)
\\
$^{1}$Canadian Institute for Theoretical Astrophysics, University of Toronto, 60 St. George Street, Toronto, ON M5S 3H8, Canada\\
$^{2}$Dunlap Institute for Astronomy and Astrophysics, University of Toronto, 50 St. George Street, Toronto, ON M5S 3H4, Canada\\
$^{3}$David A.~Dunlap Department of Astronomy and Astrophysics, University of Toronto, 50 St. George Street, Toronto, ON M5S 3H4, Canada\\
$^{4}$Center for Cosmology and Particle Physics, Department of Physics, New York University, 726 Broadway, New York, NY, 10003, USA\\
$^{5}$Institute of Theoretical Astrophysics, University of Oslo, P.O. Box 1029 Blindern, N-0315 Oslo, Norway\\
$^{6}$California Institute of Technology, 1200 E. California Blvd., Pasadena, CA 91125, USA\\
$^{7}$D\'epartement de Physique Th\'{e}orique, Universit\'{e} de Gen\`{e}ve, 24 Quai Ernest-Ansermet, CH-1211 Gen\`{e}ve 4, Switzerland\\
$^{8}$Jodrell Bank Centre for Astrophysics, Alan Turing Building, Department of Physics and Astronomy, The University of Manchester, Oxford Road, Manchester,\\M13 9PL, U.K.
}

\date{Accepted XXX. Received YYY; in original form ZZZ}

\pubyear{2022}

\begin{document}
\label{firstpage}
\pagerange{\pageref{firstpage}--\pageref{lastpage}}
\maketitle

\begin{abstract}
We present the deconvolved distribution estimator (DDE), an extension of the voxel intensity distribution (VID), in the context of future observations proposed as part of the CO Mapping Array Project (COMAP). The DDE exploits the fact that the observed VID is a convolution of correlated signal intensity distributions and uncorrelated noise or interloper intensity distributions. By deconvolving the individual VID of two observables away from their joint VID in a Fourier-space operation, the DDE suppresses sensitivity to interloper emission while maintaining sensitivity to correlated components. The DDE thus improves upon the VID by \replaced{rejecting}{reducing the relative influence of} uncorrelated noise and interloper biases, which is useful in the context of COMAP observations that observe different rotational transitions of CO from the same comoving volume in different observing frequency bands. Fisher forecasts suggest that the theoretical sensitivity in the DDE allows significant improvements in constraining power compared to either the cross power spectrum or the individual VID data, and matches the constraining power of the combination of all other one- and two-point summary statistics. Future work should further investigate the covariance and model-dependent behaviour of this novel one-point cross-correlation statistic.
\end{abstract}




\section{Introduction}
\label{sec:intro}
Line-intensity mapping (LIM; see reviews by~\citealt{LIM2017,LIM2019} and~\citealt{BernalKovetz22}) is a nascent paradigm for surveying the cosmic web at early cosmic epochs. Instead of tracing the large-scale structure (LSS) of the Universe with individual resolved galaxies, LIM proposes to use unresolved emission in atomic or molecular spectral lines across large cosmological volumes. The resulting measurements of clustering of and shot noise in the cosmological line emission should allow statistical inferences of properties of the entire population of line emitters, such as the luminosity function at both faint and bright ends, the bias with which line emission traces the underlying LSS, and so on.

While LIM thus works around challenges that conventional surveys face in target resolution and selection bias, that does not render LIM an \emph{easier} endeavour. Mitigation of systematics and contaminants is key to being able to leverage the rich statistical promise of LIM. In this respect, a LIM measurement of one spectral line on its own gives at best an incomplete and tenuous picture of cosmic evolution. Cross-correlations involving LIM experiments and other tracers of LSS will strongly reject disjoint sources of error while enabling multi-tracer astrophysics on cosmological scales to completely probe the environmental and topological factors that drive events like cosmic reionisation and galaxy assembly~\citep[see, e.g.,][]{Sun19}. Indeed, some of the earliest halo models used to forecast prospects for higher-frequency LIM experiments intended cross-correlation of carbon monoxide (CO) and ionised carbon ([C\,\textsc{ii}]) signals with 21 cm observations or other LSS tracers as a central science case~\citep{Gong11,Lidz11,Gong12,Pullen13}. Naturally cross-correlations remain a key consideration in LIM survey forecasting, design, and analysis~\citep[see, e.g.:][]{BreysseAlexandroff19,Keenan22,EXCLAIM}.

The bulk of cross-correlation forecasts focus on two-point statistics, examining the possibility of detecting cross power spectra. But two-point statistics only provide complete information about the correlation of fields if the fields are Gaussian, whereas LIM signals end up being significantly non-Gaussian at small scales due to the nonlinear nature of the process of structure formation and the resulting spatial distribution of line emitters. On the other hand, one-point statistics like the voxel intensity distribution (VID) -- i.e., the histogram of voxel intensities or temperatures across the LIM data cube -- will be sensitive to small-scale, non-Gaussian information. As previous studies have shown, one-point statistics like the VID thus critically complement two-point statistics like power spectra by breaking degeneracies inherent to the latter and thus improving the constraining power of LIM data, as shown explicitly in previous studies~\citep{Breysse17,Ihle19,Breysse22PRL}. 

An extension of the VID involves conditioning voxel intensities on external data, e.g., obtaining the VID only across voxels in the LIM data cube that overlap with a galaxy detection in an external dataset, and comparing this to the VID obtained across voxels in the LIM data cube without an overlapping galaxy. Comparing different conditional VID data rejects biases disjoint between the datasets just as cross-correlation two-point statistics do, but still retaining non-Gaussian information. \cite{Breysse19} confirmed this in an explicit simulation study of the conditional VID between 21 cm data and counts-in-cells in simulations (or more precisely, of the ratio of Fourier-transformed conditional VID data, representing an operation that deconvolves foreground biases uncorrelated with the conditioning galaxy distribution). The EXperiment for Cryogenic Large-Aperture Intensity Mapping (EXCLAIM) also recently presented conditional VID forecasts, with [C\,\textsc{ii}] intensities conditioned on galaxy counts-in-cells~\citep{EXCLAIM}. 

However, certain contexts require an extension of the VID that deals with the joint probability density function (PDF) of correlated voxel intensity components, rather than discrete conditional distributions. One such context is found in the CO Mapping Array Project (COMAP;~\citealt{Cleary22}), a dedicated single-dish LIM experiment that targets rotational transitions of CO. A key aim of COMAP in later phases is cross-correlation of $z\sim7$ CO(2--1) emission observed in the Ka band (around 30 GHz) with $z\sim7$ CO(1--0) emission observed in the Ku band (around 15 GHz). In such a study of molecular gas in the late Epoch of Reionisation (EoR), we wish to leverage a robust joint-PDF extension of the VID capable of dealing with two continuous line-intensity fields, which would strongly complement the cross power spectrum detections forecast for future COMAP phases.

In this paper, we describe the deconvolved distribution estimator (DDE;~\citealt{Breysse22DDE}). Derived from the individual and joint PDF of correlated signal intensities, the DDE is\added{ designed to be} robust to independent noise or other disjoint sources of bias in the individual VID or joint PDF, as its calculation deconvolves uncorrelated biasing distributions away from the joint PDF. We demonstrate the potential utility of the DDE in the context of the COMAP, and specifically the COMAP Expanded Reionisation Array (COMAP-ERA) concept proposed by~\cite{Breysse21}. In doing so, we will answer the following questions:

\begin{itemize}
    \item Is the DDE robust to contaminants like noise and interloper emission, as is the analytic expectation?
    \item How much could the DDE fundamentally improve constraints on $z\sim7$ CO emission in simulated COMAP-ERA observations?
\end{itemize}

We structure the paper as follows. In~\autoref{sec:motivation} we motivate and define the DDE in more mathematical detail. \autoref{sec:methods} then defines the COMAP-ERA simulations in which we propose to examine the potential of the DDE. We consider the results of these simulations in~\autoref{sec:results} before concluding in~\autoref{sec:conclusions}.

Unless otherwise stated, we assume base-10 logarithms, and a $\Lambda$CDM cosmology with parameters $\Omega_m = 0.286$, $\Omega_\Lambda = 0.714$, $\Omega_b =0.047$, $H_0=100h$\,km\,s$^{-1}$\,Mpc$^{-1}$ with $h=0.7$, $\sigma_8 =0.82$, and $n_s =0.96$, to maintain consistency with previous simulations used by~\cite{Ihle19}. Distances carry an implicit $h^{-1}$ dependence throughout, which propagates through masses (all based on virial halo masses, proportional to $h^{-1}$) and volume densities ($\propto h^3$).

\section{DDE motivation}
\label{sec:motivation}
In this section, we discuss the fundamentals of the DDE but refer the reader to~\cite{Breysse22DDE} for further theoretical grounding.

While the LIM signal traces large-scale structure and therefore has two-point correlations in comoving space as described by the power spectrum (or higher-point correlations---see, e.g.,~\citealt{BeaneLidz18}), we may model the one-point statistics of the signal approximately as a random variable following some probability distribution $\mathcal{P}_S(T)$. The variance of this signal is entirely independent of the variance of the noise in the LIM survey, which follows its own probability distribution $\mathcal{P}_N(T)$. The noise is Gaussian for an ideal spectroscopic cube from a radiometer survey; the signal is strongly non-Gaussian and asymmetric.

In the case of COMAP reionisation observations, the Ka-band and Ku-band measurements of temperatures $T_\text{Ka}$ and $T_\text{Ku}$ will contain correlated signals due to CO(2--1) and CO(1--0) emission from $z\sim7$ respectively. However, the Ka-band will also contain a lower-redshift interloper component in the form of CO(1--0) emission from $z\sim3$. This component follows its own probability distribution $\mathcal{P}_I(T_\text{Ka})$, which is independent of the signals\footnote{We expect the clustering of the emission to be largely decorrelated between the interloper and the signals, given the large interval between $z\sim3$ and $z\sim7$.} and of the radiometer noise.

The probability distribution of the sum of independent random variables is the convolution of the probability distributions of the individual variables. Therefore the observed brightness temperature, being the sum of signal and noise, will follow a distribution obtained by convolving the signal and noise distributions:
\begin{align}
    \mathcal{P}(T)=\mathcal{P}_S\ast\mathcal{P}_N(T).
\end{align}
In cases like Ka-band COMAP observations, the interloper distribution also enters the convolution:
\begin{align}
    \mathcal{P}(T_\text{Ka})=\mathcal{P}_S\ast\mathcal{P}_{N,\text{Ka}}\ast\mathcal{P}_I(T_\text{Ka}).
\end{align}
\begin{figure}
    \centering
    \begin{tikzpicture}
    \node () {\includegraphics[width=0.96\linewidth]{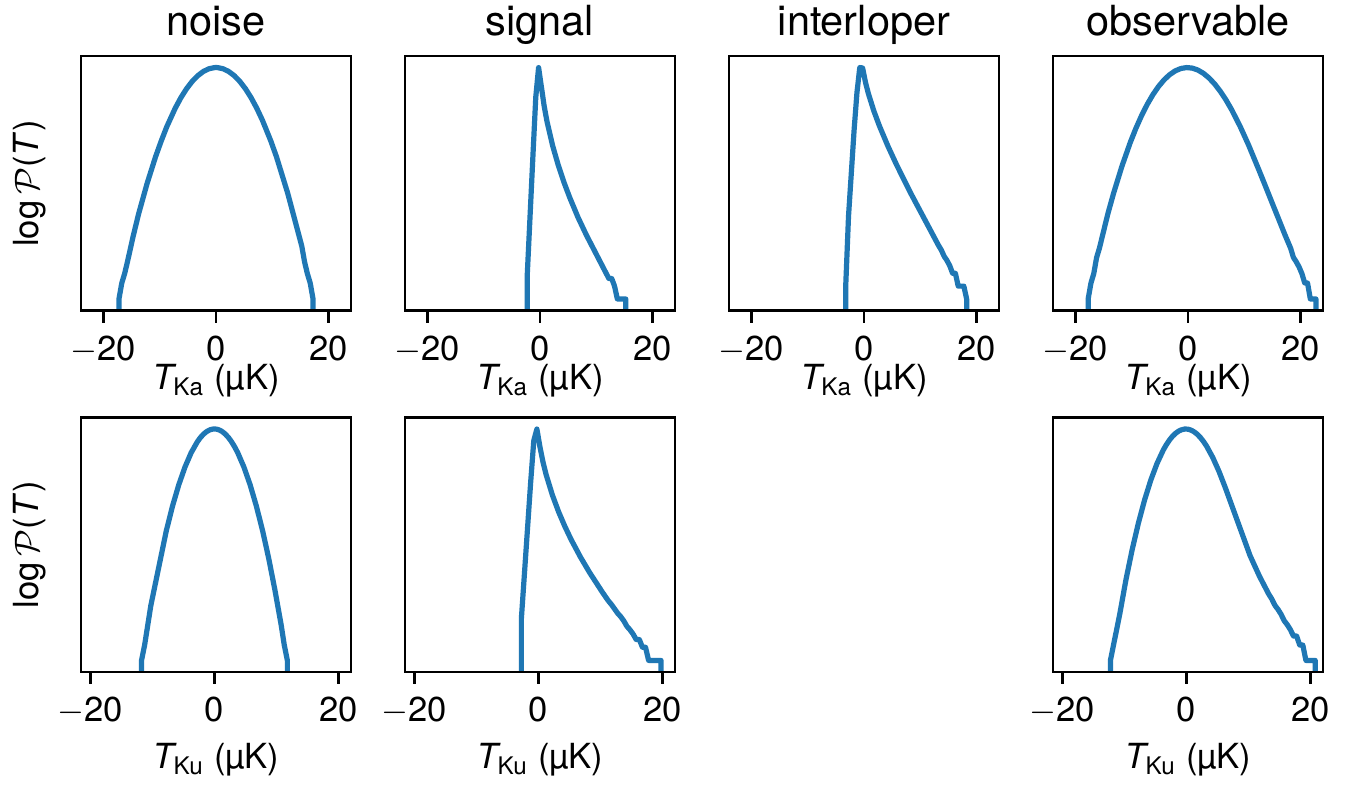}};
    \node () at (-1.8,1.386) {\Large $\ast$};
    \node () at (-1.8,-0.86555) {\Large $\ast$};
    \node () at (0.15,1.386) {\Large $\ast$};
    \node () at (2.15,1.386) {\Large $=$};
    \node () at (2.15,-0.86555) {\Large $=$};
    \end{tikzpicture}
    \caption{Illustration for Ka- and Ku-band observations (upper and lower rows respectively) of 1D distributions of noise, (mean-subtracted) signal, and (for Ka-band only) interloper voxel intensities being convolved into the observed probability distribution.}
    \label{fig:convolutions}
\end{figure}
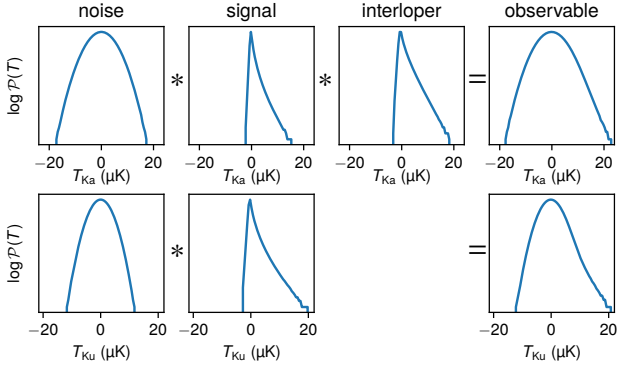

We provide a graphical representation in~\autoref{fig:convolutions}. Note that in the latter case, we can treat the interloper as an additional non-Gaussian component of `noise', meaning we treat $\mathcal{P}_{N,\text{Ka}}\ast\mathcal{P}_I$ as a total noise distribution $\mathcal{P}_N$.

Now consider the joint probability distribution between independent observations of temperatures $T_1$ and $T_2$ ($T_\text{Ka}$ and $T_\text{Ku}$ in the case of reionisation-epoch COMAP). If the noise is independent between observations and follows distributions $\mathcal{P}_{N1}(T_1)$ and $\mathcal{P}_{N2}(T_2)$, common knowledge holds that
\begin{align}
    \mathcal{P}_N(T_1,T_2)=\mathcal{P}_{N1}(T_1)\mathcal{P}_{N2}(T_2).\label{eq:jointsep}
\end{align}
On the other hand, if the signals in each observation are perfectly correlated (which is close to what simulations suggest for the low-$J$ CO lines---see~\citealt{Yang21b}), and follow effectively the same probability distribution $\mathcal{P}_S$ for both $T_1$ and $T_2$ (up to any difference in normalisation, which can be removed with rescaling), then the joint distribution would be the product of the signal distribution and a delta function:
\begin{align}
    \mathcal{P}_S(T_1,T_2)=\mathcal{P}_S(T_1)\delta(T_2-T_1).\label{eq:jointdelta}
\end{align}
As with the univariate distributions, the joint distribution of the total observed $T_1$ and $T_2$ follows the convolution of the signal and noise distributions:
\begin{align}
    \mathcal{P}(T_1,T_2)&=\mathcal{P}_S\ast\mathcal{P}_N(T_1,T_2).
\end{align}
\begin{figure}
    \centering
    \includegraphics[width=0.986\linewidth]{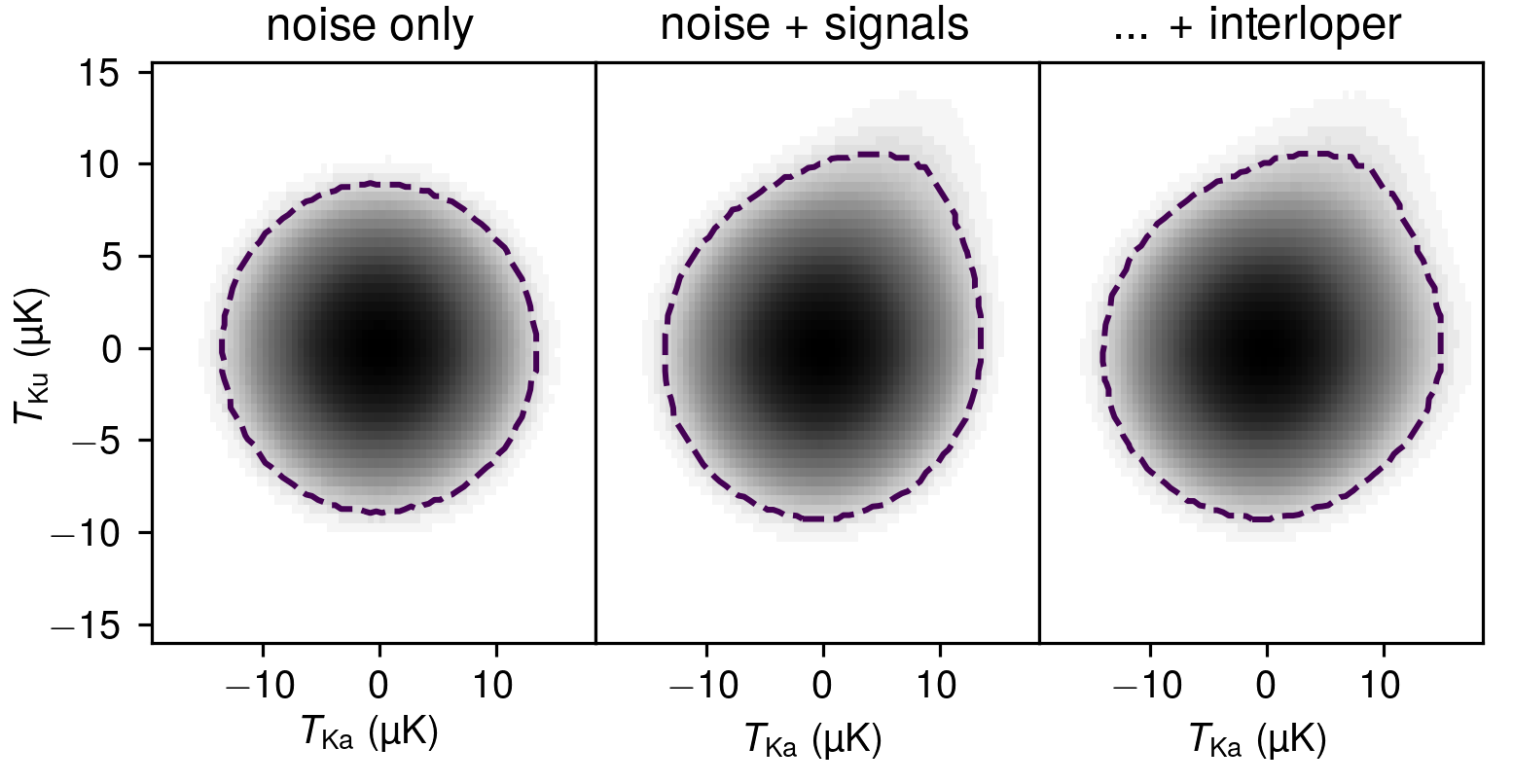}
    \caption{Bivariate probability distributions for voxel intensity given perfect white noise in both Ka- and Ku-bands (left), white noise plus correlated signals (middle), and with the additional component of a Ka-band interloper line (right). We also draw $\sim3.5\sigma$ contours (dashed) to aid visualisation of how signal and interloper intensity distributions convolve with the noise-only distribution.}
    \label{fig:2DVID}
\end{figure}

Without working explicitly through the maths, we can illustrate graphically in~\autoref{fig:2DVID} that the joint distribution of the total signal-plus-noise $T_1$ and $T_2$ in this case has a clear correlation between higher $T_1$ and higher $T_2$, a result of the signal distribution that modifies the noise distribution (which we also show) through convolution. In the case of the reionisation-epoch COMAP observation, the interloper line also modifies the distribution through convolution, but purely in the $T_\text{Ka}$ dimension and introducing no correlation with $T_\text{Ku}$. This makes sense because the interloper joint distribution would be entirely a function of $T_\text{Ka}$ and thus completely separable (and absorbable into $\mathcal{P}_{N,\text{Ka}}(T_\text{Ka})$ as before). However, it remains the case that this interloper convolution interferes with our main correlated signal. So we want to devise a statistic that is based on the joint $\mathcal{P}(T_1,T_2)$ but sensitive only to $\mathcal{P}_S$.

Now consider the Fourier transforms $\widetilde{\mathcal{P}}$ of each of the univariate and joint distributions considered here, and in particular the description of the Fourier transform of the joint distributions in terms of the Fourier transforms of the univariate distributions. We use $\widetilde{T}_1$ and $\widetilde{T}_2$ to denote the Fourier duals of the temperature variables $T_1$ and $T_2$.

The Fourier transform of a convolution of functions is the product of the Fourier transforms of the individual functions, univariate or otherwise:
\begin{align}
    \widetilde{\mathcal{P}}(\widetilde{T}_1)&=\widetilde{\mathcal{P}}_S(\widetilde{T}_1)\widetilde{\mathcal{P}}_{N1}(\widetilde{T}_1)\label{eq:PT1}\\
    \widetilde{\mathcal{P}}(\widetilde{T}_2)&=\widetilde{\mathcal{P}}_S(\widetilde{T}_2)\widetilde{\mathcal{P}}_{N2}(\widetilde{T}_2)\label{eq:PT2}\\
    \widetilde{\mathcal{P}}(\widetilde{T}_1,\widetilde{T}_2)&=\widetilde{\mathcal{P}}_S(\widetilde{T}_1,\widetilde{T}_2)\widetilde{\mathcal{P}}_{N}(\widetilde{T}_1,\widetilde{T}_2)\label{eq:PT12}
\end{align}

For separable joint distributions like the noise distribution described in~\autoref{eq:jointsep}, the Fourier transform is simply the product of the Fourier transforms of the separated distributions:
\begin{align}
    \widetilde{\mathcal{P}}_N(\widetilde{T}_1,\widetilde{T}_2)=\widetilde{\mathcal{P}}_{N1}(\widetilde{T}_1)\widetilde{\mathcal{P}}_{N2}(\widetilde{T}_2).\label{eq:PN12}
\end{align}
In the specific case of the perfectly correlated joint signal described in~\autoref{eq:jointdelta}, working through the Fourier transform results in
\begin{align}
    \widetilde{\mathcal{P}}_S(\widetilde{T}_1,\widetilde{T}_2)=\widetilde{\mathcal{P}}_S(\widetilde{T}_1+\widetilde{T}_2).\label{eq:PS12}
\end{align}

In analogue to normalised time-series cross-correlation or covariance (e.g., Pearson's bivariate correlation coefficient), we define the DDE here as a normalised measure of correlation between variables $T_1$ and $T_2$ via their Fourier duals $\widetilde{T}_1$ and $\widetilde{T}_2$:
\begin{align}
    \mathcal{D}\equiv\frac{\widetilde{\mathcal{P}}(\widetilde{T}_1,\widetilde{T}_2)}{\widetilde{\mathcal{P}}(\widetilde{T}_1)\widetilde{\mathcal{P}}(\widetilde{T}_2)}-1.\label{eq:defDDE}
\end{align}

We can then ask what $\mathcal{D}$ is for $T_1$ and $T_2$ when including both signal and noise (with any separable interlopers presumed as fully described by one of the univariate Fourier-transformed noise distributions) as described above. Substituting Equations~\ref{eq:PT1}--\ref{eq:PS12} into~\autoref{eq:defDDE}, we find
\begin{align}
    \mathcal{D}&=\frac{\widetilde{\mathcal{P}}_S(\widetilde{T}_1,\widetilde{T}_2)\widetilde{\mathcal{P}}_N(\widetilde{T}_1,\widetilde{T}_2)}{\widetilde{\mathcal{P}}_S(\widetilde{T}_1)\widetilde{\mathcal{P}}_{N1}(\widetilde{T}_1)\widetilde{\mathcal{P}}_S(\widetilde{T}_2)\widetilde{\mathcal{P}}_{N2}(\widetilde{T}_2)}-1\\&=\frac{\widetilde{\mathcal{P}}_S(\widetilde{T}_1+\widetilde{T}_2)\widetilde{\mathcal{P}}_{N1}(\widetilde{T}_1)\widetilde{\mathcal{P}}_{N2}(\widetilde{T}_2)}{\widetilde{\mathcal{P}}_S(\widetilde{T}_1)\widetilde{\mathcal{P}}_{N1}(\widetilde{T}_1)\widetilde{\mathcal{P}}_S(\widetilde{T}_2)\widetilde{\mathcal{P}}_{N2}(\widetilde{T}_2)}-1\\&=\frac{\widetilde{\mathcal{P}}_S(\widetilde{T}_1+\widetilde{T}_2)}{\widetilde{\mathcal{P}}_S(\widetilde{T}_1)\widetilde{\mathcal{P}}_S(\widetilde{T}_2)}-1.
\end{align}
The upshot is that at least in this idealised case, $\mathcal{D}$ depends solely on the one-point statistics of the signal common to both observations, provided that any noise (including interlopers) is independent between the observations. Had there been no common signal -- i.e., had we had entirely unrelated signal distributions $\mathcal{P}_S(T_1)$ and $\mathcal{P}_S(T_2)$ in the two measurements -- we would have obtained $\mathcal{D}=0$, so any deviation from $\mathcal{D}=0$ indicates the presence of a correlated signal in both distributions. (An anti-correlated signal would similarly result in deviation from $\mathcal{D}=0$, as we could have defined the joint distribution in~\autoref{eq:jointdelta} with $\delta(T_1+T_2)$ and ended up with $\widetilde{P}_S(\widetilde{T}_1-\widetilde{T}_2)$ as the joint signal distribution in~\autoref{eq:PS12}.)

In practice, we must deal with discrete histograms, and discrete fast Fourier transforms rather than continuous Fourier transforms (with corresponding normalisation factors necessary as part of the computational implementation depending on convention). However, the basic idea is sound and we proceed to apply it to simulated COMAP observations.

\section{Demonstration: COMAP simulation}
\label{sec:methods}

We need to define three key ingredients in order to simulate COMAP survey volumes: a model for line emission (\autoref{sec:linemodel}), parameters for the COMAP-ERA survey (\autoref{sec:survey}), and large numbers of approximate cosmological simulations (\autoref{sec:sims}) to which we can apply the first two ingredients.
\subsection{Fiducial CO model}
\label{sec:linemodel}
Of the CO models considered by~\cite{Breysse21}, we use the adaptation of the model of~\cite{Li16} by~\cite{mmIME-ACA} at all redshifts. The basic idea behind the model of~\cite{Li16} is to connect halo mass $M_h$ and redshift $z$ to an average star-formation rate based on the empirical model of~\cite{Behroozi13b,Behroozi13a}, which is assumed to be related linearly to IR luminosity by a conversion factor of $10^{10}\,L_\odot\,(M_\odot\text{ yr}^{-1})^{-1}$. Empirical power-law fits to data then relate IR luminosity to CO($J\to J-1$) luminosity:
\begin{equation}
    \log{\frac{L_\text{IR}}{L_\odot}} = \alpha_J\left(\log{\frac{L_{\text{CO},J}}{L_\odot}}+4.31\right)+\beta_J
\end{equation}
At this step, \cite{mmIME-ACA} diverge from the fiducial model of~\cite{Li16} out of a need (which we share) to model multiple rotational transitions of CO. The \cite{Li16}--\cite{mmIME-ACA} model uses the work of~\cite{Kamenetzky16}, which obtains $\alpha_1=1.27$, $\beta_1=-1.0$, $\alpha_2=1.11$, and $\beta_2=-0.6$ based on a local sample of galaxies. The model also includes random log-normal scatter to mimic variations in galaxy properties, both around the average star-formation rate for fixed halo mass and redshift, and around the average CO luminosity for fixed star-formation rate. For each property there is an independent scatter introduced with a standard deviation of 0.3 in units of dex, but we calculate the star-formation rate for each halo only once and use it to inform both CO line luminosities. This results in imperfectly but non-negligibly correlated CO(1--0) and CO(2--1) line luminosities, although perhaps less correlated than semi-analytic models of CO emission suggest (see, e.g.,~\citealt{Yang21b}).

We will want to constrain $\alpha_1$, $\beta_1$, $\alpha_2$, and $\beta_2$ at $z\sim7$. In addition, most summary statistics end up being chiefly sensitive to the luminosity-weighted average CO temperature--bias product $\avg{Tb}_J$ for each CO line, defined as
\begin{equation}
    \avg{Tb}_J = \frac{c^3(1+z)}{8\pi k_B\nu_{\text{rest},J}^3 H(z)}\int dM_h\,\frac{dn}{dM_h}\,L_{\text{CO},J}(M_h,z)\,b(M_h).\label{eq:Tbdef}
\end{equation}
This calculation requires: a model relation $b(M_h)$ for the linear halo bias, the scaling with which the number density contrast of halos of virial mass $M_h$ traces the underlying matter density contrast; a halo mass function $dn/dM_h$ to describe halo number densities; the rest frequency of the CO($J\to J-1$) line $\nu_{\text{rest},J}\approx115.27\cdot J$ GHz; the Hubble parameter $H(z)$; and standard physical constants, namely the speed of light $c$ and the Boltzmann constant $k_B$. In this context, we will use the $b(M_h)$ model of~\cite{Tinker10}, and a corrected form of the~\cite{Tinker08} halo mass function given by~\cite{Behroozi13b} (which we will also use below in~\autoref{sec:sims}). Given this, the fiducial model results in $\avg{Tb}_1=3.37$\,\textmu K and $\avg{Tb}_2=2.51$\,\textmu K.

We will refer to inferences both of $\{\alpha_J,\beta_J\}$ and of $\avg{Tb}_J$, with the latter calculable from the former by obtaining credibility intervals for $\avg{Tb}_J$ implied by the $L_{\text{CO},J}(M_h)$ relations found from the credibility intervals for $\alpha_J$ and $\beta_J$. In addition, we will not assume that the IR--CO relation is the same at $z\sim3$ as it is at $z\sim7$, and will allow $\alpha_{1,z\sim3}$ and $\beta_{1,z\sim3}$ to be inferred separately from $\alpha_1$ and $\beta_1$ (with the latter pair taken to be referring to $z\sim7$ parameters).

\subsection{COMAP parameters}
\label{sec:survey}
We base our CO survey parameters on the COMAP-ERA parameters considered by~\cite{Breysse21}. The survey is assumed to span three fields of $2^\circ\times2^\circ$ each, covering frequencies of 13--17 GHz in the Ku band and 26--34 GHz in the Ka band. These volumes are discretised as a grid of $N_\text{pix}\times N_\text{ch}=(30\times30)\times512$ voxels, each spanning four arcminutes in both angular dimensions to match the beam size (assumed to be 4.5 arcminutes across the Ka band, and 3.9 arcminutes across the Ku band), and 7.8125 MHz or 15.625 MHz respectively in the Ku or Ka band.

Based on the nominal system temperature $T_\text{sys}$ of each receiver, the spectrometer count\footnote{\cite{Breysse21} use the term `feeds'; we attempt to disambiguate further. Each Ka-band dish will host 19 single-polarisation feeds and thus 19 spectrometers; each Ku-band dish will host 19 dual-polarisation feeds and thus 38 spectrometers.} per dish $N_\text{spd}$, the science channel bandwidth $\delta\nu$, the pixel count $N_\text{pix}$ per field, and the number of dish-hours per field $N_\text{dish}t_\text{obs}$, we can calculate the radiometer noise per voxel as
\begin{equation}
    \sigma_N = \frac{T_\text{sys}}{\sqrt{\delta\nu N_\text{spd}N_\text{dish}t_\text{obs}/N_\text{pix}}}.
\end{equation}
We show the resulting values for both Ku- and Ka-band observations in~\autoref{tab:comapparams} alongside a summary of other key parameters.

\begin{table}
    \centering
    \begin{tabular}{ccc}
         \hline Parameter & \multicolumn{2}{c}{Values for:} \\
         & Ku band & Ka band\\\hline
         Frequency coverage (GHz) & 13--17 & 26--34\\
         Nominal system temperature $T_\text{sys}$ (K) & 20 & 44\\
         Spectrometer count per dish $N_\text{spd}$ & 38 & 19\\
         Science channelisation $\delta\nu$ (MHz) & 7.8125 & 15.625\\
         Dish-hours per field $N_\text{dish}t_\text{obs}$ (hr) & 57000 & 110000\\
         Noise per voxel $\sigma_N$ (\textmu K) & 2.55 & 3.85
         \\\hline
    \end{tabular}
    \caption{Key parameters for COMAP-ERA observations either taken from~\protect\cite{Breysse21} or newly assumed for this work.}
    \label{tab:comapparams}
\end{table}

\subsection{Peak-patch simulations}
\label{sec:sims}
As was the case for the VID at the time of writing of~\cite{Ihle19}, the covariance matrix of our observables has aspects not well-understood in a purely analytic formalism, and we will use a large number of simulations to estimate the covariance matrix numerically. We use the peak-patch method~\citep{Stein19} to obtain large numbers of dark matter simulations and halo catalogues from independent sets of initial conditions.

\cite{Ihle19} used 161 independent peak-patch lightcone simulations spanning a comoving volume of $L_\text{box}^3=(1140\,$Mpc$)^3$ with a resolution of $N_\text{cells}=4096^3$. As the lightcone extent was equivalent to $9.6^\circ\times9.6^\circ$ in transverse dimensions and 26--34 GHz in CO(1--0) observing frequency,~\cite{Ihle19} were able to split each lightcone into many subfields to simulate thousands of semi-independent COMAP observations at $z\sim3$. We will still be able to use these same simulations to generate realisations of the $z\sim3$ CO(1--0) interloper emission in the Ka-band mock data cube. However, we require additional peak-patch simulations at high redshift, and with slightly better mass resolution. Given the filtering scales used in the peak-patch method to find matched density peaks, the minimum resolvable halo mass is given by
\begin{align}
    M_\text{h,min-res}&=\frac{4}{3}\pi\Omega_m\rho_c\left(\frac{2L_\text{box}}{N_\text{cells}^{1/3}}\right)^3\\&=9.3\times10^{12}\Omega_mh^2\frac{(L_\text{box}/\text{Mpc})^3}{N_\text{cells}}M_\odot\\&=1.3\times10^{12}\frac{(L_\text{box}/\text{Mpc})^3}{N_\text{cells}}M_\odot.
\end{align}
For the peak-patch simulations used by~\cite{Ihle19} this is $2.8\times10^{10}\,M_\odot$, which is within 10\% of the minimum halo mass quoted by~\cite{Ihle19} of $2.5\times10^{10}\,M_\odot$ (noting that~\cite{Ihle19} correct the halo masses obtained from the peak-patch method via abundance matching to the~\cite{Tinker08} HMF). While this is sufficient for the $z\sim3$ CO signal, the $z\sim7$ CO signal will have greater contribution from the lower-mass halo population and we want to make sure the statistics for that population are as correct as possible down to at least several times lower mass. At the same time, we still want to obtain sufficiently large volumes to be divisible into many semi-independent $z\sim7$ COMAP observations. We therefore choose a box size of $L_\text{box}=960$\,Mpc and $N_\text{cells}=5640^3$, so that $M_\text{h,min-res}=6.4\times10^9\,M_\odot$.

The original $z\sim3$ simulations used the SciNet General Purpose Cluster (GPC;~\citealt{SciNetGPC}), before its decommissioning in April 2018. Each run used 900 seconds of compute time on 2048 Intel Xeon E5540 cores (or 256 nodes), using $\simeq2.4$ TB of RAM (roughly half the available RAM, as most GPC nodes had 16 GB RAM). Our new $z\sim7$ simulations use the successor to the GPC, the Niagara cluster~\citep{SciNetNiagara}, whose nodes use a mix of Intel Xeon 6148 and 6248 CPU cores at the time of this work. These simulations also use $\approx900$ seconds runtime per realisation with 1880 cores (or 47 nodes), but with a peak memory footprint of $\simeq7.5$ TB (around 90\% of the available RAM) owing to the higher resolution. Just as the 161 $z\sim3$ realisations from~\cite{Ihle19} took only $\sim82000$ CPU hours, we are able to generate 270 realisations after only $\approx127000$ core-hours. We estimate (as did~\citealt{Ihle19} and~\citealt{Stein19}) that this is three orders of magnitude faster than an equivalent N-body simulation.

The resulting halo catalogues span $z=5.8$--$7.9$ and $6^\circ\times6^\circ$. As with the $z\sim3$ halo catalogues, we correct peak-patch halo masses through abundance matching to the HMF of~\cite{Tinker08}, but in this case with high-redshift corrections from Appendix G of~\cite{Behroozi13a}. We use these corrections only in mass-correcting our new simulations as they principally apply at $z>3$.

By dividing the $z\sim3$ simulations into 16 sub-patches each spanning $2^\circ\times2^\circ$, and the $z\sim7$ simulations into 9 sub-patches each spanning the same, we obtain 2430 semi-independent $2^\circ\times2^\circ$ sky realisations, which we believe should be sufficient for a reasonably high-quality estimation of the covariance in the context of this early investigatory work.

We use \texttt{limlam\_mocker}\footnote{\url{https://github.com/georgestein/limlam\_mocker}} to assign halo luminosities and create simulated brightness temperature cubes for Ku-band CO(1--0) and Ka-band CO(2--1), as well as the Ka-band CO(1--0) interloper. We also include the line broadening model described by~\cite{linewidths}. We also apply a high-pass transfer function to mimic the effect of the COMAP pipeline filtering out large-scale angular and frequency modes as described by~\cite{Foss21} and~\cite{Ihle21}. However, while we adapt the form suggested by~\cite{COMAPESV}, we adjust the exponents slightly to reflect that we project the data cube onto comoving space at $z\sim7$, rather than at $z\sim3$ as for COMAP Pathfinder analyses:
\begin{equation}
    \mathcal{T}_\text{hp}(k_\perp,k_\parallel) = \frac{1}{(1+e^{5-138\text{\,Mpc}\cdot k_\perp})(1+e^{5-144\text{\,Mpc}\cdot k_\parallel})},
\end{equation}
where $k_\parallel$ is the line-of-sight component of the wavevector $\mathbfit{k}$, and $k_\perp$ is the magnitude of the transverse part of $\mathbfit{k}$ (so $k_\perp^2+k_\parallel^2=k^2$).

We calculate the following summary statistics for each data cube:
\begin{itemize}
    \item the Ku-band auto power spectrum, with 101 $k$-bins of width $\Delta k=0.02$, centred at values ranging from 0.01\,Mpc$^{-1}$ to 2.09\,Mpc$^{-1}$;
    \item the Ka-band auto power spectrum, with the same $k$-bins used for the Ku-band;
    \item the cross power spectrum between the Ku- and Ka-band data cubes, again using the same $k$-bins;
    \item the Ku-band `auto' VID, using a histogram across $0.5$\,\textmu K-wide bins spanning $T_\text{Ka}\in(-36,36)$\,\textmu K;
    \item the Ka-band `auto' VID, using a histogram across $0.5$\,\textmu K-wide bins spanning $T_\text{Ka}\in(-48,48)$\,\textmu K;
    \item and the DDE, based on fast Fourier transforms of the individual and joint VID with the same binning of $T_\text{Ku}$ and $T_\text{Ka}$ as used for the individual VID calculations.
\end{itemize}

\section{Results}
\label{sec:results}
\begin{figure}
    \centering
    \includegraphics[width=0.986\linewidth]{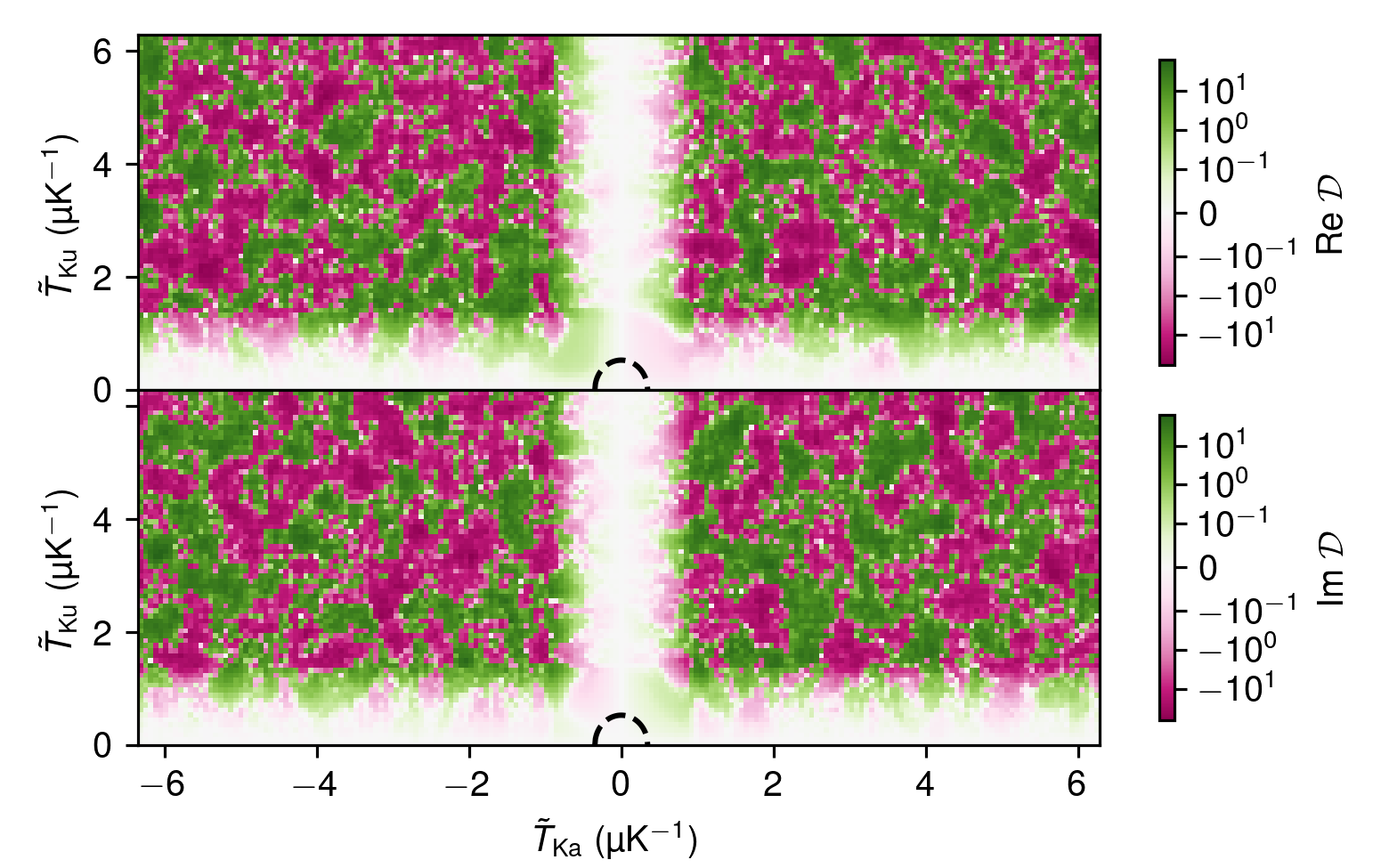}
    \caption{Illustration of the DDE real and imaginary parts (upper and lower sub-panels respectively) when all signal, noise, and interloper cubes are accounted for, using the median across all realisations. The dashed ellipse indicates the low-variance cut in both $\widetilde{T}$ dimensions described in the main text as a function of the noise distribution.}
    \label{fig:inspection2d}
\end{figure}
\begin{figure*}
\centering\includegraphics[width=0.493\linewidth]{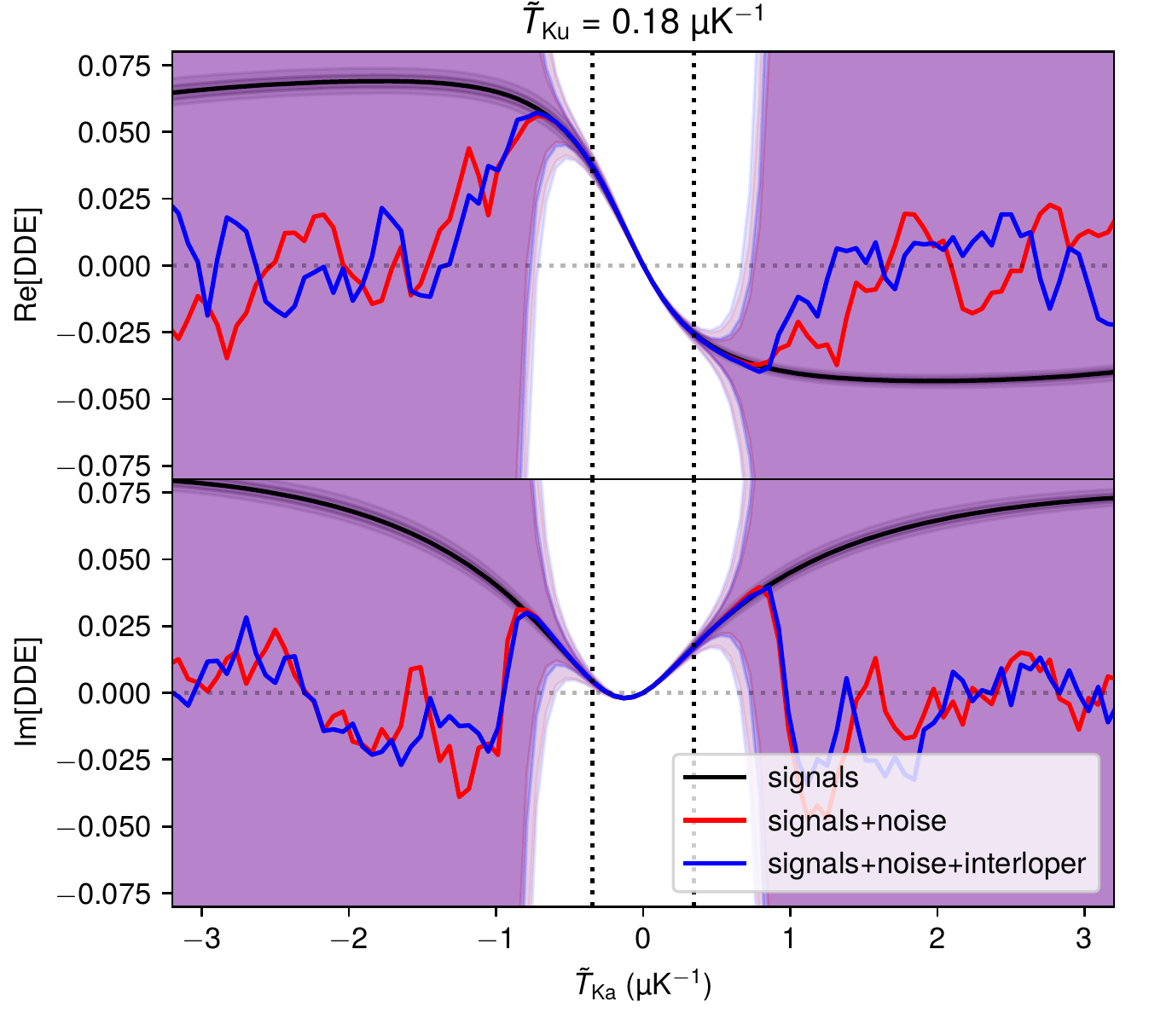}
\includegraphics[width=0.493\linewidth]{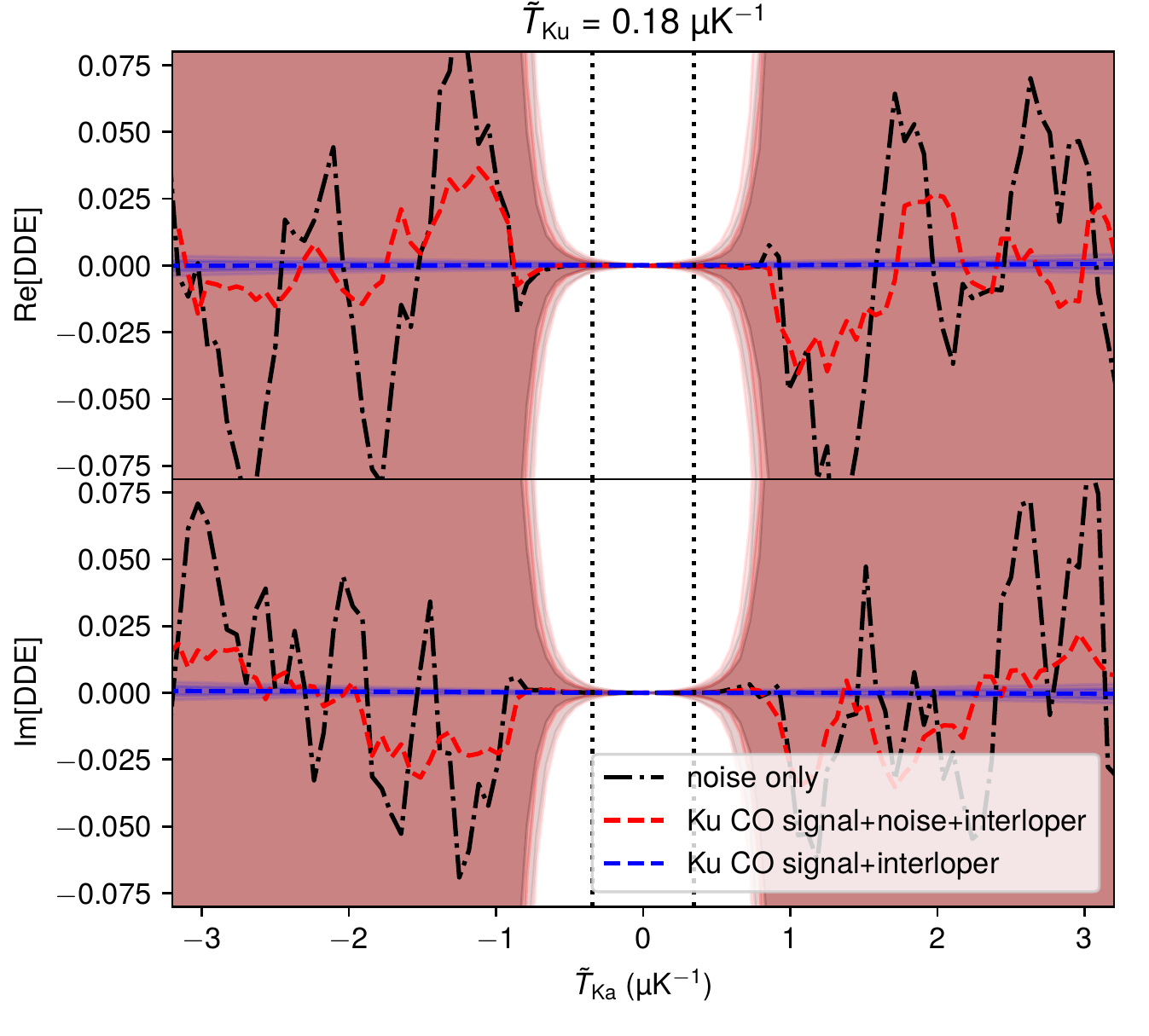}
    \caption{Slices of the DDE and low-variance cut (now shown as dotted vertical lines), again showing real and imaginary parts separately (upper and lower sub-panels respectively, within each panel). Across the left and right panels, we show how the DDE changes if certain simulation components are omitted, in particular whether the Ka-band observation contains the CO(2--1) signal correlated with the Ku-band CO(1--0) signal (left panel) or whether the Ka-band observation contains no such correlated signal (right panel). Dark (light) shaded areas around each curve indicate 68\% (95\%) sample intervals, with the presence of noise significantly increasing interval widths.}
    \label{fig:inspection}
\end{figure*}
\begin{figure}
\centering\includegraphics[width=0.986\linewidth]{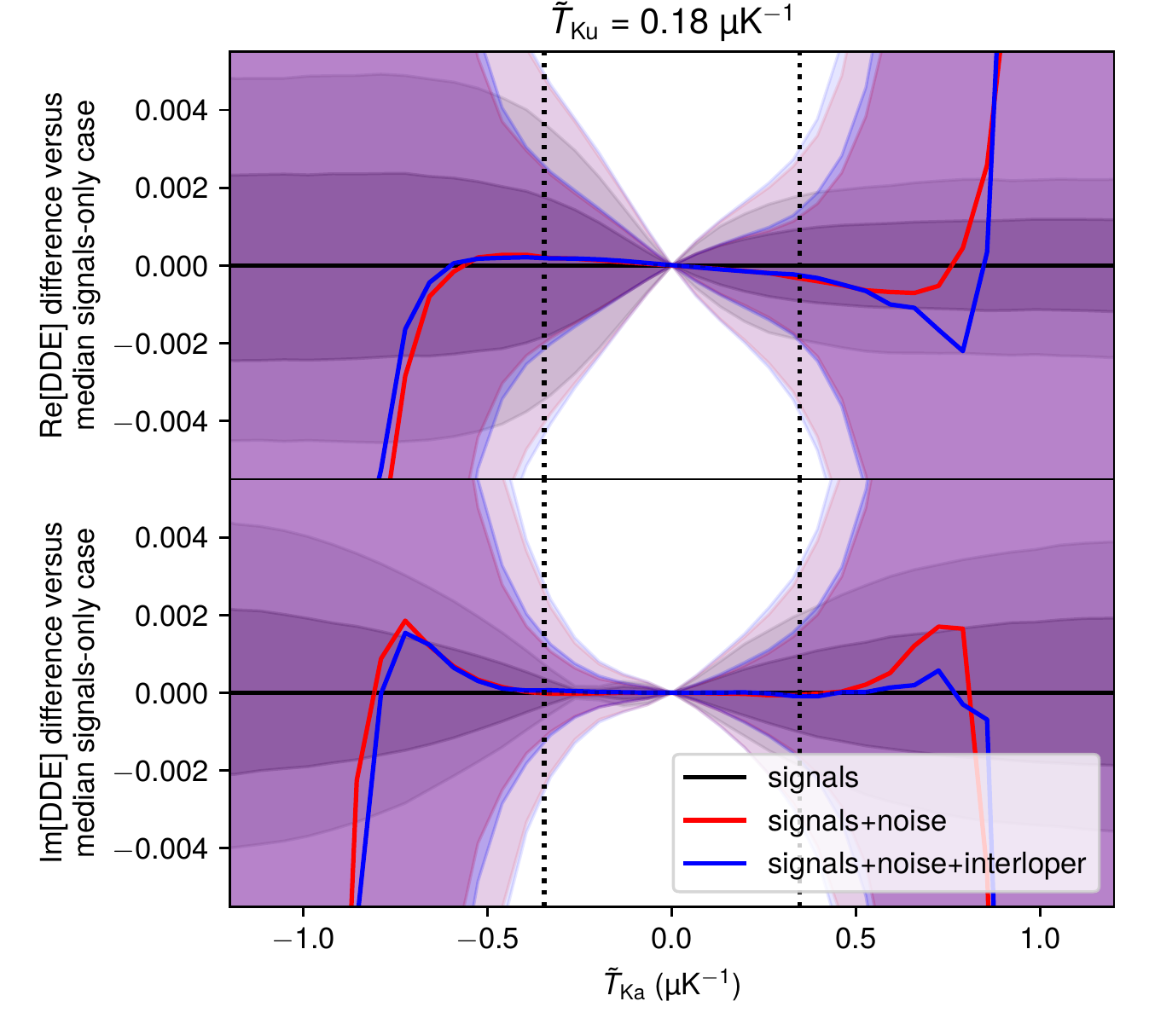}
    \caption{Same as the left panel of~\autoref{fig:inspection}, but now showing residuals relative to the median obtained from realisations with only the correlated signals. Again, dark (light) shaded areas around each curve indicate 68\% (95\%) sample intervals.}
    \label{fig:inspectionres}
\end{figure}

\begin{figure*}
    \centering
    \includegraphics[width=0.495\linewidth,clip=True,trim=3mm 0 3mm 0]{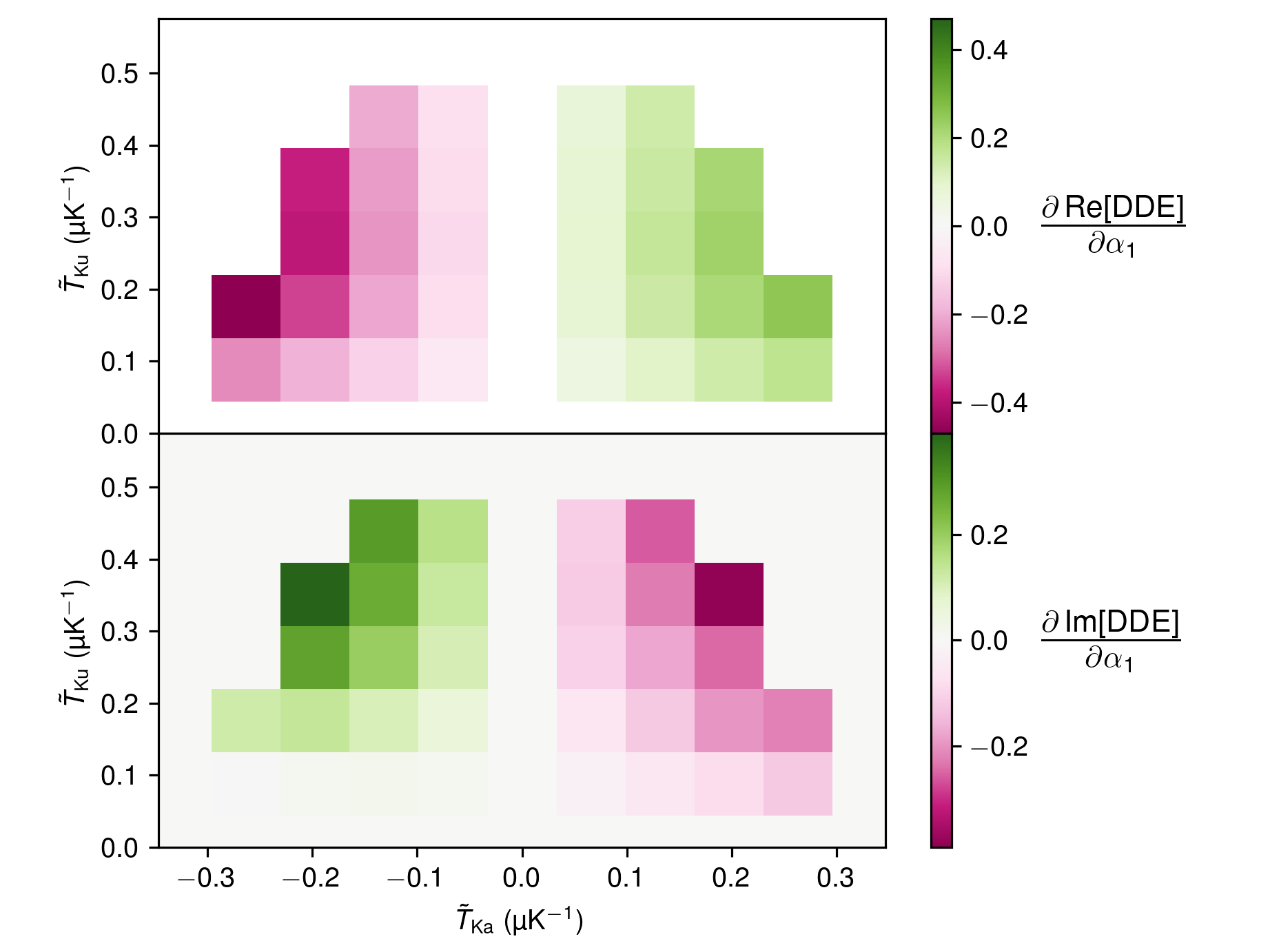}\includegraphics[width=0.495\linewidth,clip=True,trim=3mm 0 3mm 0]{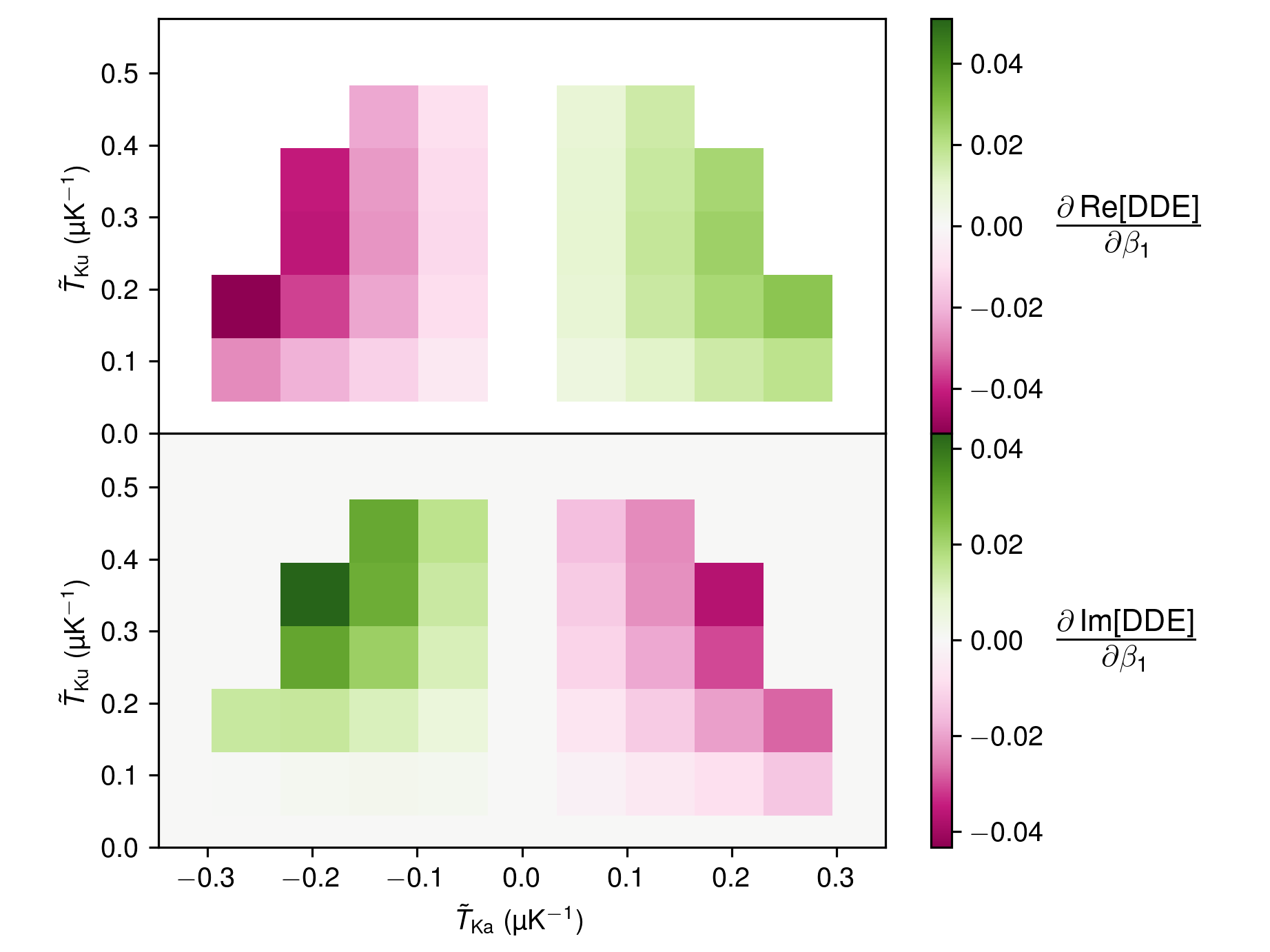}
    
    \includegraphics[width=0.495\linewidth,clip=True,trim=3mm 0 3mm 0]{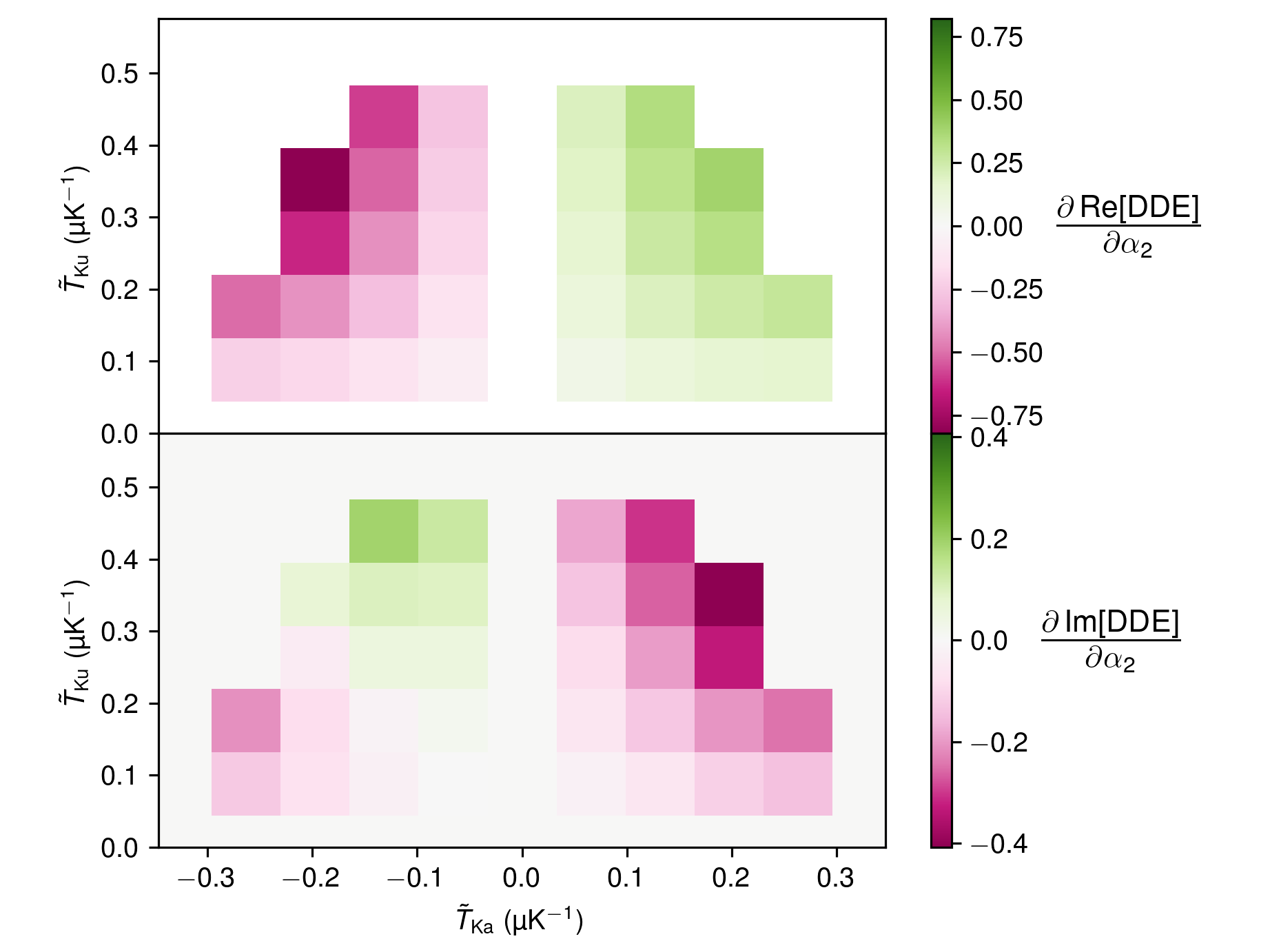}\includegraphics[width=0.495\linewidth,clip=True,trim=3mm 0 3mm 0]{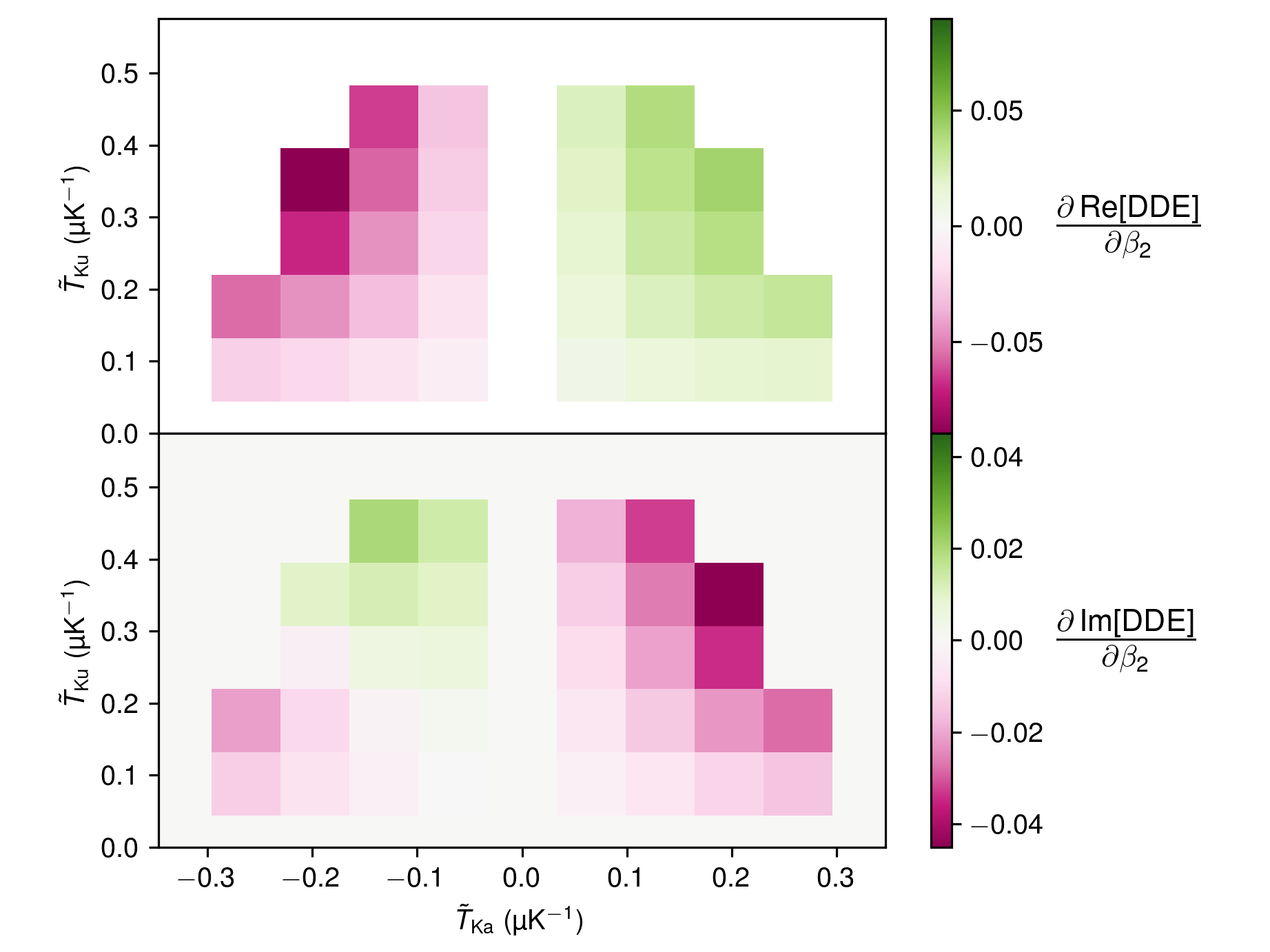}
    
    \includegraphics[width=0.495\linewidth,clip=True,trim=3mm 0 3mm 0]{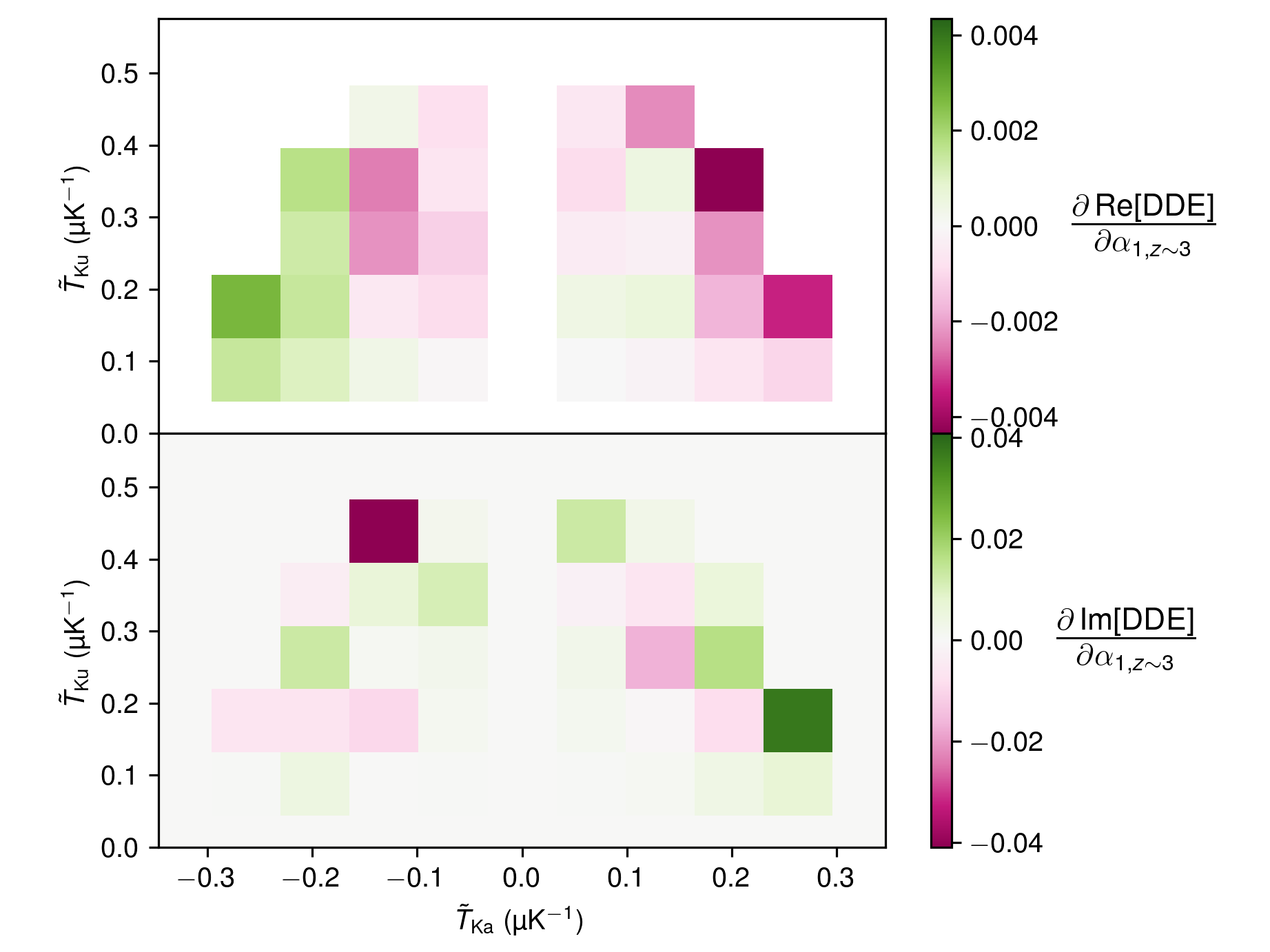}\includegraphics[width=0.495\linewidth,clip=True,trim=3mm 0 3mm 0]{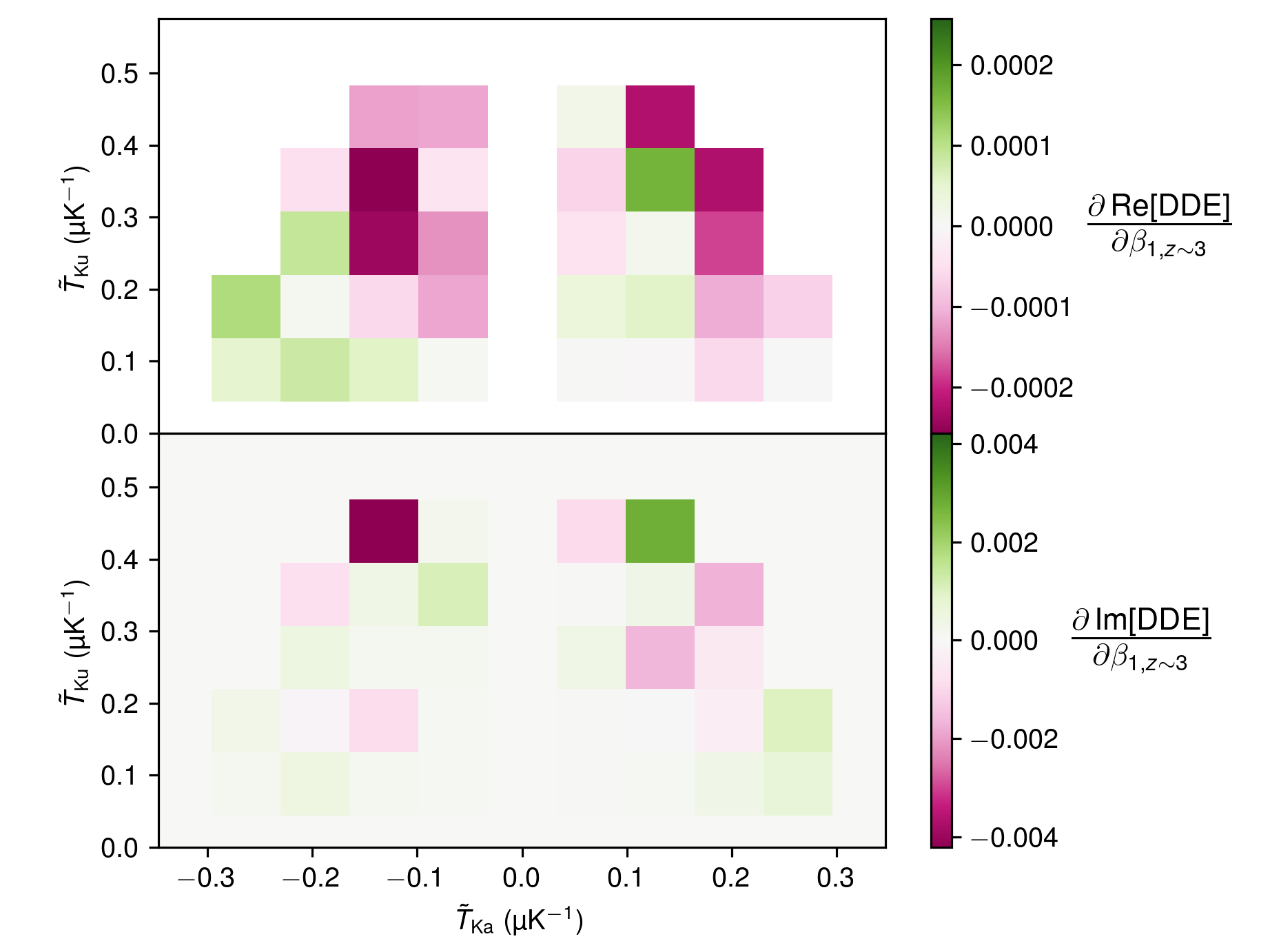}
    \caption{Partial derivatives of the real and imaginary parts of the DDE with respect to model parameters, as indicated by the label next to the colour bars. Compare in particular the partial derivatives of the imaginary part of the DDE for negative $\widetilde{T}_\text{Ka}$ against $\alpha_1$ or $\beta_1$ versus those against $\alpha_2$ or $\beta_2$.}
    \label{fig:weirdness}
\end{figure*}

\subsection{Visual inspection}
\label{sec:inspect}
We show the typical DDE for our simulated observations in~\autoref{fig:inspection2d}, and show slices of the DDE with different components of the simulation included or excluded in~\autoref{fig:inspection}. Since the noise distribution is the most visible component, we suggest that the most usable modes of the DDE (i.e., those least affected by noise) will be larger-scale modes than the Fourier scale set by the instrumental noise. We set the cutoff to correspond to when the half-wavelength of the mode equals the full width at half-maximum of the noise distribution:
\begin{align}
    \widetilde{T}_\text{Ka}&<\frac{\pi}{2.355\sigma_{N,\text{Ka}}};\\
    \widetilde{T}_\text{Ku}&<\frac{\pi}{2.355\sigma_{N,\text{Ku}}}.
\end{align}
The values at $\widetilde{T}=0$ are also unusable as they are always equal to $0+0i$, and we only choose the modes with $\widetilde{T}_\text{Ku}>0$ as the other half of the Fourier modes in the DDE are completely dependent. For our $0.5$\,\textmu K-wide bins spanning $T_\text{Ka}\in(-48,48)$\,\textmu K and $T_\text{Ku}\in(-36,36)$\,\textmu K, this in principle leaves 32 usable DDE bins\footnote{By contrast, using the present binning scheme, only two DDE bins would be usable with data from the COMAP-EoR experiment preceding COMAP-ERA, also conceptualised by~\cite{Breysse21}. However, an actual COMAP-EoR analysis would tailor the VID binning appropriately to the wider noise distribution, so would have more than just two DDE bins available.}, each with a real and imaginary part.

Looking at the increase in DDE variance introduced by noise in~\autoref{fig:inspectionres}, it is qualitatively true that the noise variance is much lower inside the $\widetilde{T}$ cutoff than outside it. It is possible that the cut is somewhat conservative and DDE data at slightly larger values of $\widetilde{T}$ could be usable, but considering only the bins within the cutoff will be sufficient for the exploratory purposes of this work.

Of interest is whether what we expect based on the continuous distribution formalism of~\autoref{sec:motivation} holds in these numerical simulations. With our fiducial binning the outcome is as good as can be expected, at least by eye. But overall outlook here is mixed due to some numerical instability. In a preliminary version of this analysis that used coarser 2\,\textmu K bins, introducing noise did in fact affect the DDE by dampening its deviation from $0+0i$, contrary to our expectation. On the other hand, once noise was introduced in addition to the correlated signals, introducing the Ka-band interloper emission did not appear to further affect the DDE even in this case. Furthermore, in that case also, when there was no correlated signal in the Ku-band simulated observation, taking the DDE between the Ka- and Ku-band cubes resulted in a value of $0+0i$ throughout, as expected.

All this suggests \emph{some} robustness of the DDE (or rather the fact of its deviation from $0+0i$) modulo choice of binning, and with sufficiently fine histogramming\footnote{Note in particular that both the noise per voxel $\sigma_N$ and the CO $\avg{Tb}_J$ are around $3$\,\textmu K, so the histogramming likely needs to over-sample this typical fluctuation magnitude by a factor of several, which is the case with 0.5\,\textmu K bins but not 2\,\textmu K bins.} the robustness matches theoretical expectations \replaced{perfectly}{much better}. Furthermore, even before we consider astrophysical interpretation, the level of variance in~\autoref{fig:inspection} suggests a very high-significance detection of correlated distributions is achievable with COMAP-ERA.

The important aspect for parameter inference, however, is the covariance (which we will estimate numerically) and the interplay of the covariance with the partial derivatives of the DDE with respect to each of the high-redshift CO parameters $\alpha_J$ and $\beta_J$ as well as the nuisance low-redshift CO parameters $\alpha_{1,z\sim3}$ and $\beta_{1,z\sim3}$. We estimate this numerically from a smaller set of 96 simulations, for which we repeat calculations of all summary statistics while shifting one parameter away from its fiducial value. $\alpha_J$ and $\alpha_{1,z\sim3}$ are allowed to shift $\pm0.02$ away from their fiducial values; $\beta_J$ and $\beta_{1,z\sim3}$ shift a larger $\pm0.2$ away as the signals are not as sensitive to the same absolute change in those parameters.

We show the resulting central difference quotient estimates in~\autoref{fig:weirdness}. We note with interest that the imaginary part of the DDE reacts in a visibly qualitatively different way to shifts in $\alpha_2$ compared to shifts in $\alpha_1$; the same applies to $\beta_J$.\added{ That is, changing CO(2--1) parameters alters the shapes of the individual and joint VID in a different way to changing CO(1--0) parameters. Different temperature-space offsets introduced in the distribution shapes would correspond to different phases introduced in the DDE and thus the different partial derivatives for the real and imaginary parts of the DDE that we see in~\autoref{fig:weirdness} against $\alpha_2$ and $\beta_2$ versus against $\alpha_1$ and $\beta_1$.} This will become important in a moment in considering the astrophysical constraints possible with this measure, as we will discuss in~\autoref{sec:fisherres}.

Note also that we find non-zero partial derivatives of the DDE against nuisance parameters. As the range of values indicated in~\autoref{fig:weirdness} demonstrates, the DDE is still significantly less sensitive to $\alpha_{1,z\sim3}$ and $\beta_{1,z\sim3}$ compared to $\alpha_J$ and $\beta_J$, and it is likely that some if not much of what we find is numerical noise. Nonetheless, given the consistency of some qualitative trends between the partial derivatives with respect to $\alpha_{1,z\sim3}$ and $\beta_{1,z\sim3}$, some real trends may well exist\added{. The effect of discrete VID binning cannot be discounted here, given our earlier observation that coarser binning reduced robustness of the DDE against noise. It is similarly possible that even finer binning than considered here would reduce any of these apparently real trends related to interloper parameters}, which we leave for future, more extensive work to examine in greater detail. Either way, \added{in the present work }we will end up forecasting non-zero constraining power on these \added{$z\sim3$ CO }parameters, although the DDE should constrain $\alpha_J$ and $\beta_J$ somewhat more tightly.

\subsection{Fisher analysis}
\label{sec:fisherres}
Our simulations allow us to numerically estimate the covariance matrix given the fiducial parameters and the full suite of 2430 simulations described in~\autoref{sec:sims}. The previously discussed smaller set of 96 simulations at various points in the local neighbourhood is sufficient to then estimate derivatives of observables with respect to the $z\sim7$ $\alpha_J$ and $\beta_J$ as well as $\alpha_{1,z\sim3}$ and $\beta_{1,z\sim3}$. This in turn allows us to run a Fisher analysis to estimate the constraining power on the $z\sim7$ $\alpha_J$ and $\beta_J$, with $\alpha_{1,z\sim3}$ and $\beta_{1,z\sim3}$ still allowed to vary as nuisance parameters to allow for the possibility of redshift evolution (although the fiducial values are the same as $\alpha_1$ and $\beta_1$ at $z\sim7$).

Note that of the 64 real variables associated with the usable DDE bins (32 real parts and 32 imaginary parts), only 29 are usable after masking variables until no pair of variables has a correlation coefficient above 0.98. The masked DDE then has an acceptably stable numerically estimated covariance matrix\footnote{Despite high correlations between the remaining DDE bins, the masked DDE covariance matrix has a condition number of $\sim10^7$, which is acceptable for 64-bit float arithmetic. When combining the masked DDE with other variables, we rescale all DDE values by a factor of $10^6$ to decrease the range of variances and improve the condition of the covariance matrix to be well beyond what is required given 64-bit float precision.}. We graphically show the correlation coefficient matrix (which has the same structure as the covariance matrix, except with variances for each observable normalised away) in~\hyperref[sec:corrcoef]{Appendix~\ref{sec:corrcoef}}.

The numerically estimated covariance matrix is calculated relative to one simulated $2^\circ\times2^\circ$ patch, so we divide it by three to reflect the fact that the full COMAP-ERA survey is of three independent fields. Once we have the full COMAP-ERA covariance matrix $\mathbfss{C}_{ab}$ for a vector of combined or individual observables $\mathbfit{O}_a$, the Fisher matrix in the basis of parameters $\{\lambda_i\}=\{\alpha_1,\alpha_2,\beta_1,\beta_2,\alpha_{1,z\sim3},\beta_{1,z\sim3}\}$ is
\begin{equation}
    \mathbfss{F}_{ij} = \sum_{a,b}\frac{\partial{\mathbfit{O}}_{a}}{\partial\lambda_i}\mathbfss{C}^{-1}_{ab}\frac{\partial{\mathbfit{O}}_{b}}{\partial\lambda_j}.\label{eq:fisher}
\end{equation}
Inverting this will yield the parameter covariance matrix, allowing us to calculate expected parameter constraints under the assumption of Gaussian covariances throughout. As some non-Gaussianity is present in our signals and variances, and as we numerically estimate all covariances and derivatives informing the Fisher forecast, we only trust the resulting parameter constraints on a qualitative level. In other words, the exact shapes and widths of the posterior distributions should not be considered robust predictions, but relative strengths of constraints between different sets of observables should be credible.

\subsubsection{Cross-statistics only}

\begin{figure}
    \centering
    \includegraphics[width=0.96\linewidth]{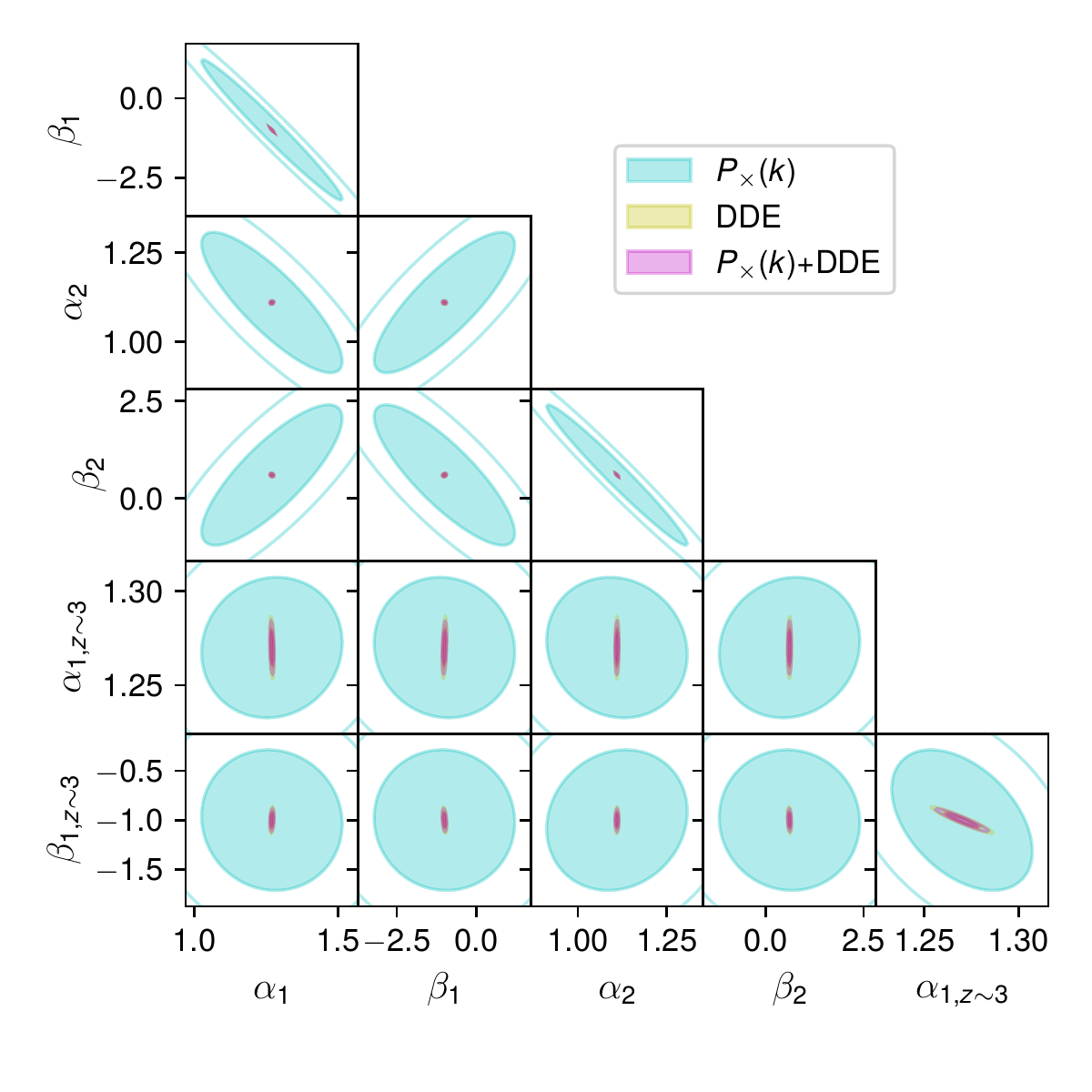}
    
    \includegraphics[width=0.96\linewidth,clip=True,trim=0 3mm 0 8mm]{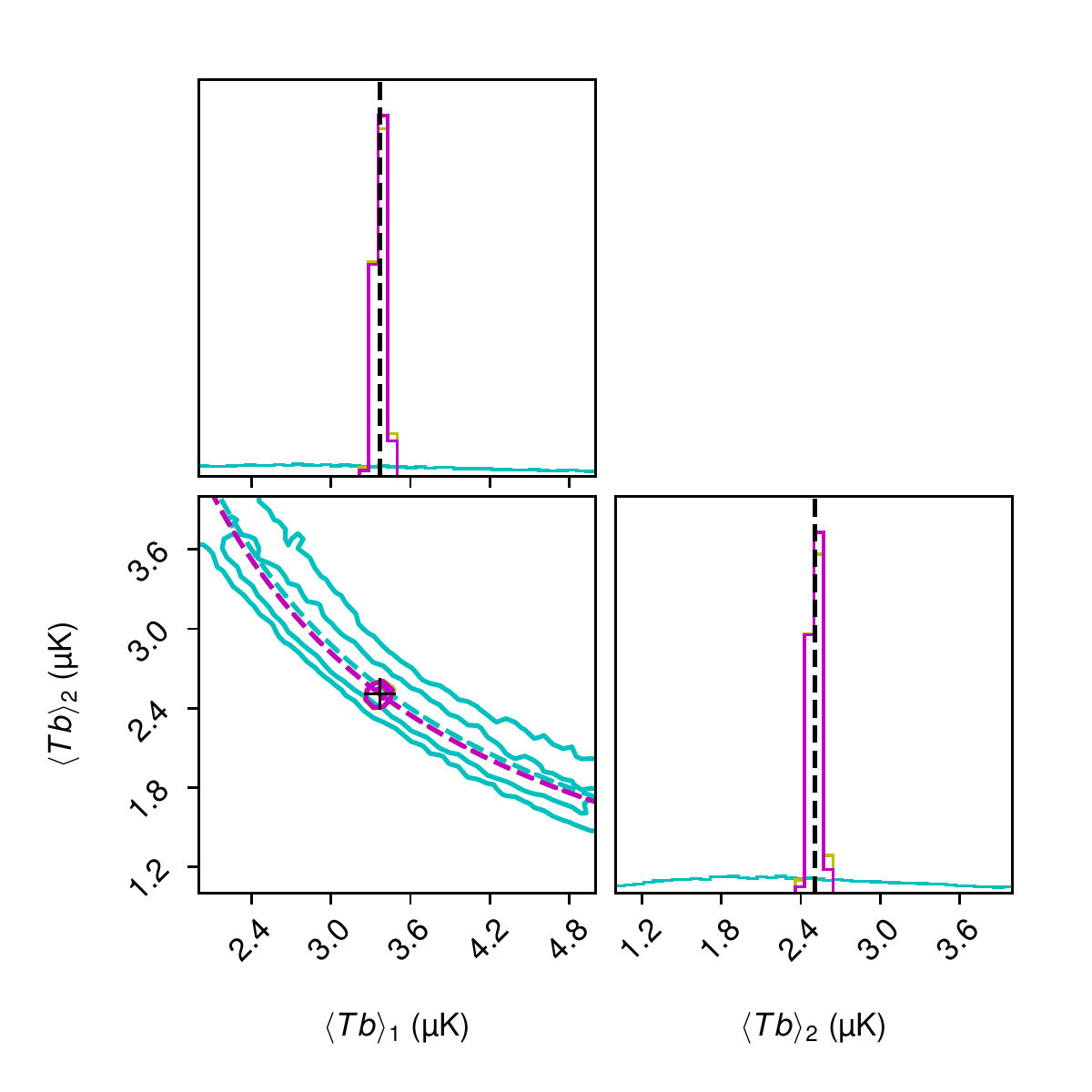}
    \caption{Corner plots of CO model parameter constraints (upper portion of plot) and constraints on $\avg{Tb}_J$ (lower portion of plot) based on the cross power spectrum alone (cyan), the masked DDE alone (yellow), and the combination of the two (magenta). (Here, the last is effectively the same as the DDE-only case, obscuring the DDE-only contours as a result. Figures~\ref{fig:fisher_allVID} and~\ref{fig:fisher_all} show the DDE-derived constraints in greater detail.) The filled (unfilled) ellipses or inner (outer) contours in each triangle plot indicate 68\% (95\%) credibility intervals. Dashed lines in the $\avg{Tb}_1$--$\avg{Tb}_2$ contour plot mark out $\avg{Tb}_1\avg{Tb}_2=8.5$\,\textmu K$^2$, which is consistent with constraints from all summary statistics.}
    \label{fig:fisher1}
\end{figure}

We show expected resulting constraints on $\alpha_J$ and $\beta_J$, as well as on the nuisance parameters $\alpha_{1,z\sim3}$ and $\beta_{1,z\sim3}$, from the Ka--Ku cross spectrum $P_\times(k)$ and the DDE in the upper panels of~\autoref{fig:fisher1}. The expected achievable constraining power of the DDE is \emph{extremely} high, far beyond that expected for the cross power spectrum. The fundamental sensitivity of the DDE appears to allow for very strong parameter constraints by itself, without use of individual VID data or power spectra. This may be a peculiar aspect of the reionisation-epoch CO signal or even our model of it, and may not necessarily hold in other LIM contexts.

We also note that the DDE does actually appear sensitive to $\alpha_{1,z\sim3}$ and $\beta_{1,z\sim3}$. However, the sensitivity to those parameters is \replaced{several}{two to three} times poorer than to any of the \added{corresponding }$z\sim7$ $\alpha_J$ and $\beta_J$ parameters.\added{ We tabulate the Fisher forecast errors for these parameters in~\hyperref[sec:fishererrors]{Appendix~\ref{sec:fishererrors}}, not only for $P_\times(k)$ and/or the DDE but for all statistics considered in this work.}

Using random draws from the parameter confidence ellipses of~\autoref{fig:fisher1}, we can translate expected constraints on $\alpha_J$ and $\beta_J$ into constraints on $\avg{Tb}_J$, shown in the lower panels of~\autoref{fig:fisher1}. With the Ka--Ku cross-spectrum alone, the degeneracy between $\alpha_J$ or between $\beta_J$ represents an inability to constrain $\avg{Tb}_J$ independently, as the cross-spectrum is proportional to their product $\avg{Tb}_1\avg{Tb}_2$. The cross spectrum constraint on this product is $\avg{Tb}_1\avg{Tb}_2=8.6_{-0.6}^{+1.1}$ (68\% interval).

However, as previously noted the DDE is somehow sensitive to CO(1--0) and CO(2--1) in different ways to the cross-spectrum, thus allowing for constraints of $\avg{Tb}_1$ and $\avg{Tb}_2$ separately. This is of extreme interest as it would allow us to simultaneously probe excitation through line ratios and molecular gas content at high redshift. The DDE constraint on the product of the two temperature--bias products is also still tighter than with the cross spectrum, at $\avg{Tb}_1\avg{Tb}_2=8.5\pm0.2$ (68\% interval).

\subsubsection{One-point statistics only}
\begin{figure}
    \centering
    \includegraphics[width=0.96\linewidth]{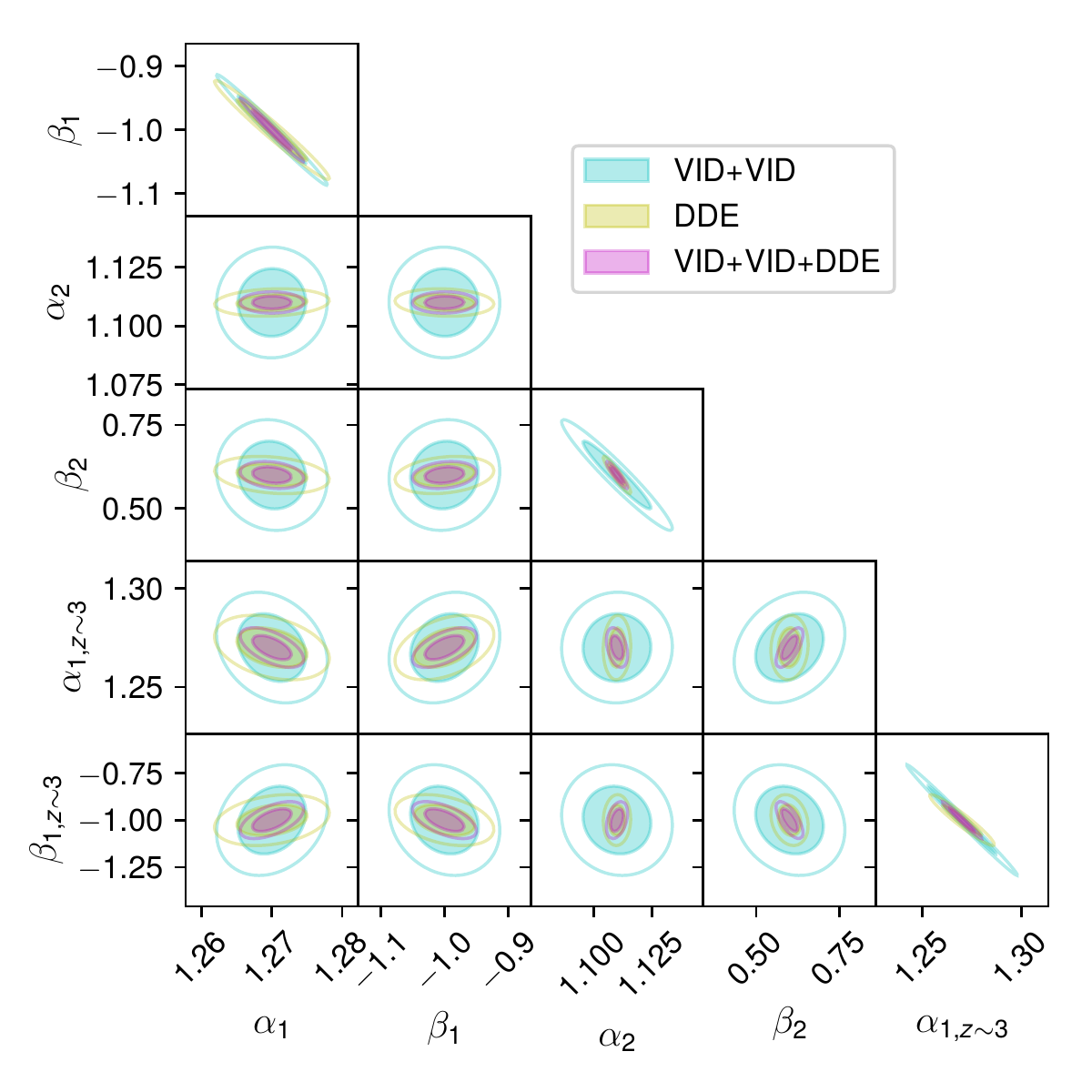}
    
    \includegraphics[width=0.96\linewidth,clip=True,trim=0 3mm 0 8mm]{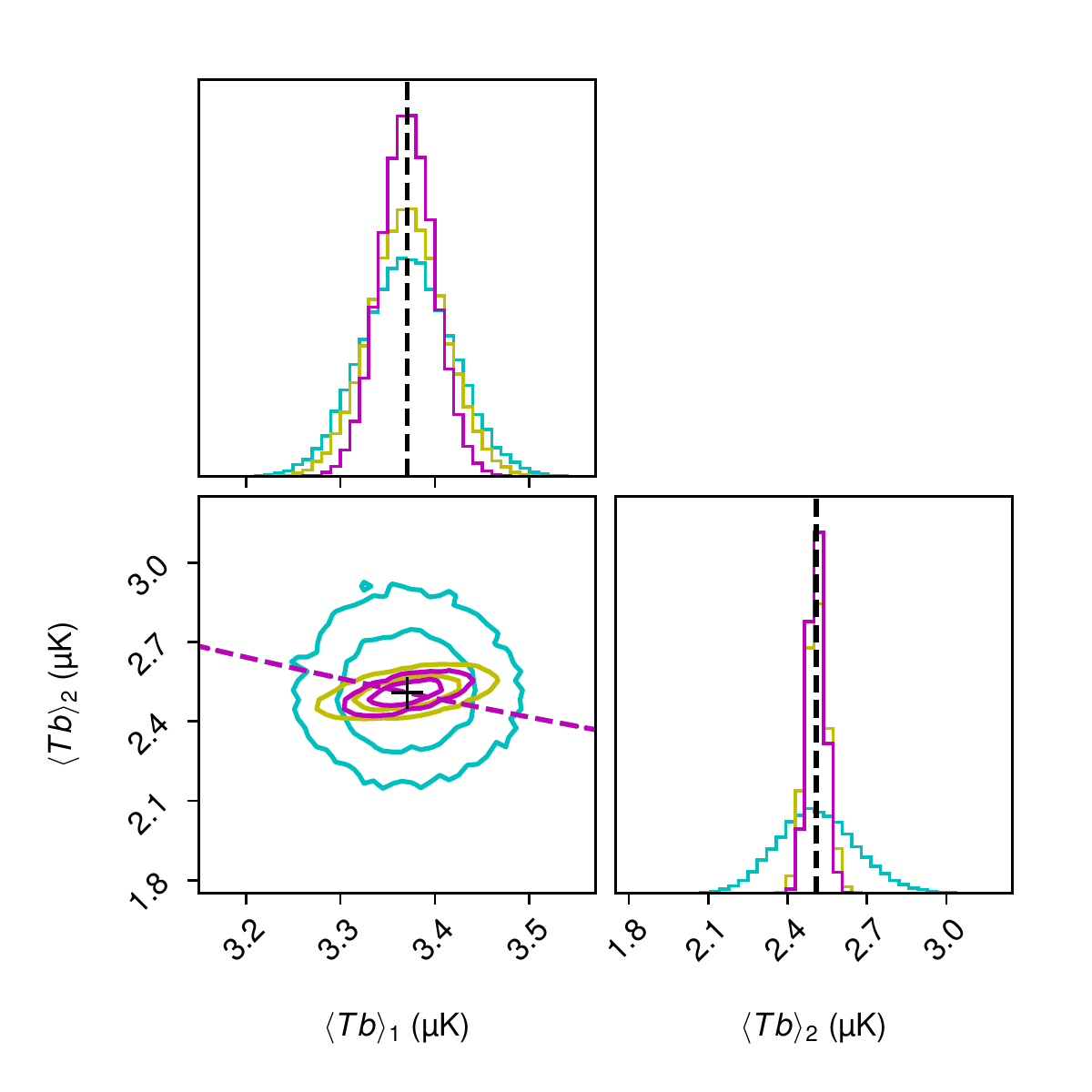}
    \caption{Same as~\autoref{fig:fisher1}, but now comparing the DDE against the combination of the individual Ku- and Ka-band VID data (possibly also combined with the DDE) instead of the cross power spectrum.}
    \label{fig:fisher_allVID}
\end{figure}

We repeat the exercise but replace the cross-spectrum with the VID from both bands (excising the first 50 and last 30 bins so that we only consider bins with significant variance). The constraints shown in~\autoref{fig:fisher_allVID} suggest that the DDE significantly collapses the parameter space volume surrounding the $z\sim7$ CO(2--1) parameters $\alpha_2$ and $\beta_2$ (as well as their nuisance $z\sim3$ CO(1--0) counterparts). This is sensible as the Ka-band VID has significant interloper emission and is thus less sensitive to the $z\sim7$ signal than either the Ku-band VID (which has no such interloper to contend with) or the DDE (which largely rejects the interloper emission by design).

The $\avg{Tb}_1$--$\avg{Tb}_2$ constraints shown in~\autoref{fig:fisher_allVID} reflect this, as both the VID and the DDE constrain $\avg{Tb}_1$ and $\avg{Tb}_2$ with little degeneracy but the DDE significantly narrows the plausible range of $\avg{Tb}_2$. This in turn also happens to significantly narrow the plausible range of $\avg{Tb}_1\avg{Tb}_2$, from $\avg{Tb}_1\avg{Tb}_2=8.5\pm0.5$ (68\% interval) for the auto VID data alone to the previously mentioned $\avg{Tb}_1\avg{Tb}_2=8.5\pm0.2$ with the DDE.

Thus, in principle, the fundamental sensitivity of COMAP-ERA would allow strong constraints on reionisation-epoch CO from one-point statistics alone. This does assume the data cubes are sufficiently clean and free of systematics and foregrounds, but again this would affect the `auto' VID far more than the DDE, which should again be significantly more robust (even if not perfectly insensitive) to such factors by design.

\subsubsection{The kitchen sink}
We now consider throwing everything we have at the problem---all auto- and cross-spectra as well as both VIDs---and examine how the DDE might \emph{still} improve constraints.

The combination of every observable other than the DDE finally matches the DDE-based constraint of $\avg{Tb}_1\avg{Tb}_2=8.5\pm0.2$ (68\% interval). With the DDE and all-but-DDE constraints on similar levels, the DDE contributes non-negligibly to the `kitchen-sink' constraint using all observables, which is $\avg{Tb}_1\avg{Tb}_2=8.5\pm0.1$ (68\% interval).

\begin{figure}
    \centering
    \includegraphics[width=0.96\linewidth]{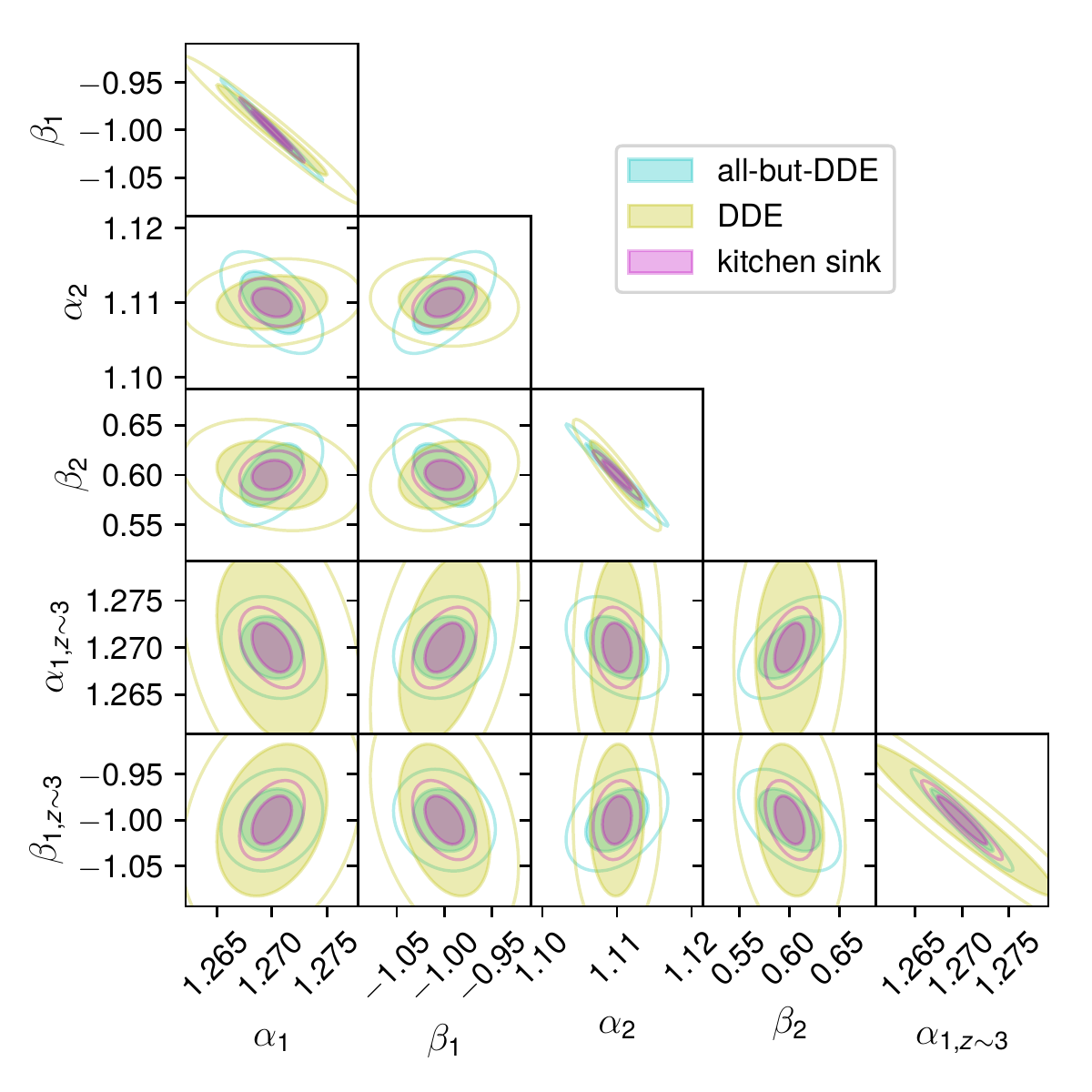}
    
    \includegraphics[width=0.96\linewidth,clip=True,trim=0 3mm 0 8mm]{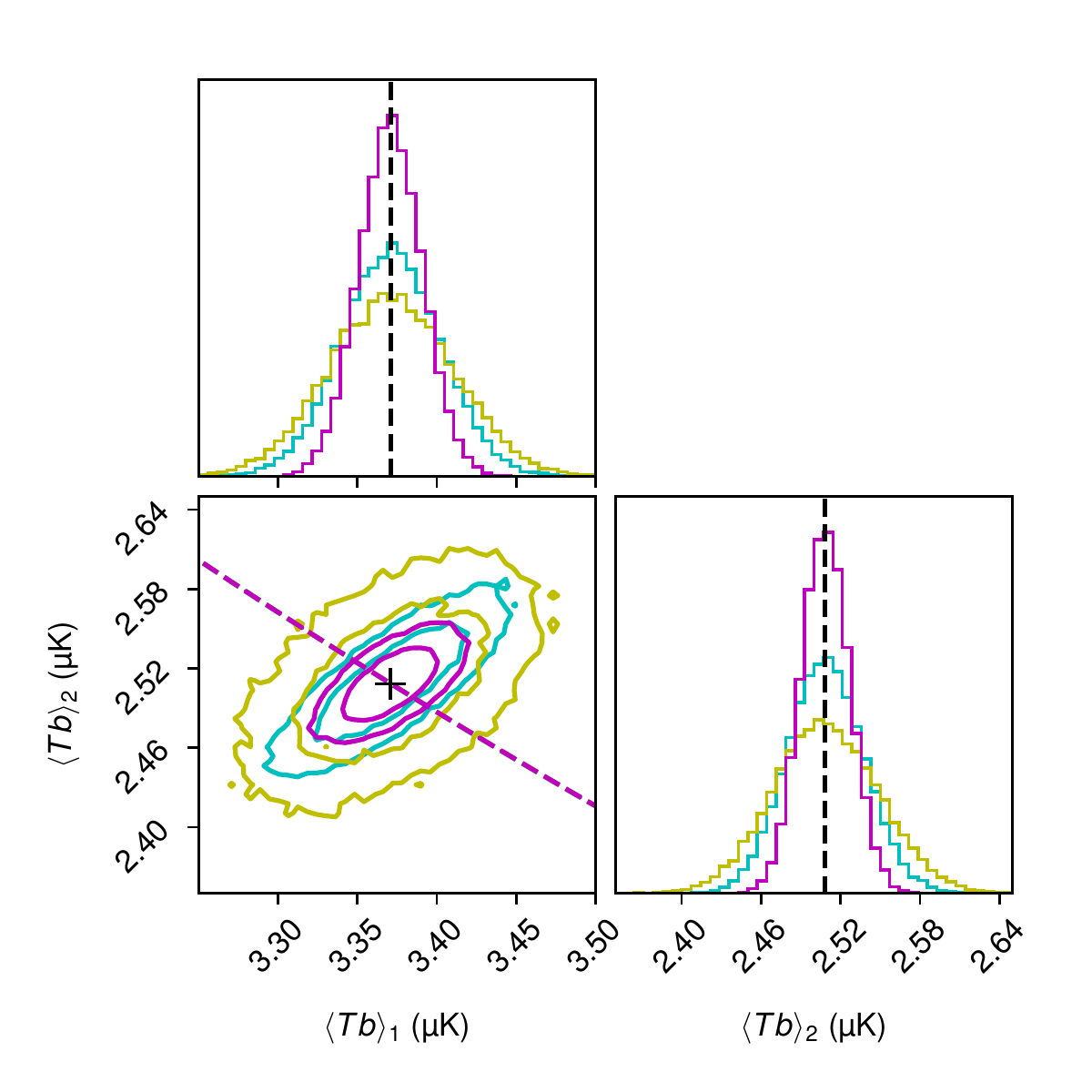}
    \caption{Same as~\autoref{fig:fisher1} or~\autoref{fig:fisher_allVID}, but now comparing the DDE against not just the cross power spectrum or just one-point statistics, but the `all-but-DDE' combination of all summary statistics from COMAP-ERA data other than the DDE (possibly also combined with the DDE in a `kitchen sink' scenario).}
    \label{fig:fisher_all}
\end{figure}

Neither $\{\alpha_J,\beta_J\}$ nor $\avg{Tb}_J$ necessarily provide an intuitive picture of physical conditions at reionisation, so we may also consider how the parameter posteriors translate to other quantities. The example we will consider briefly here is the ratio of the average CO(2--1) and CO(1--0) line brightness temperatures:
\begin{equation}
    \frac{\avg{T}_2}{\avg{T}_1} = \frac{\nu_{\text{rest},1}^3}{\nu_{\text{rest},2}^3}\frac{\int dM_h\,(dn/dM_h)\,L_{\text{CO},2}(M_h,z)}{\int dM_h\,(dn/dM_h)\,L_{\text{CO},1}(M_h,z)},\label{eq:Tratiodef}
\end{equation}
with $z=6.68$ at the midpoints of the COMAP observing frequency bands. The ratio of these temperatures is an approximate global measure of excitation of the CO rotational transitions overall and thus characterises the temperatures and densities of the interstellar medium hosting the CO gas.

The value of this ratio recovered with the DDE is $0.712\pm0.010$, and while the `kitchen-sink' constraint of $0.712\pm0.004$ is significantly better, the addition of the DDE does very marginally narrow it from the all-but-DDE constraint of $0.712\pm0.005$. Of course, improvements from the DDE would be more significant if some subset of auto power spectra or individual VID data were considered unreliable.

\section{Conclusions}
\label{sec:conclusions}
We have now answered the questions set out by the \hyperref[sec:intro]{Introduction} of this paper:

\begin{itemize}
    \item \emph{Is the DDE robust to contaminants like noise and interloper emission, as is the analytic expectation?} For appropriately fine choices of binning, we are able to define a region of Fourier-dual temperature space unaffected by noise where the DDE is an unbiased measurement of the correlation between the shapes of the distributions of two correlated variables.\added{ While we do not demonstrate perfect insensitivity to interloper emission, our analysis does show relative insensitivity of the DDE to interloper CO(1--0) parameters compared to the equivalent $z\sim7$ CO parameters.}
    \item \emph{How much could the DDE fundamentally improve constraints on $z\sim7$ CO emission in simulated COMAP-ERA observations?} The DDE potentially significantly improves constraints on the $z\sim7$ CO line model, with (for example) constraints on $\avg{Tb}_1\avg{Tb}_2$ becoming 3--5 times tighter than either the cross power spectrum alone or `auto' one-point statistics alone. Unlike the cross power spectrum, the DDE actually appears able to separately constrain $\avg{Tb}_1$ and $\avg{Tb}_2$. Even in comparison to the combination of all other available summary statistics, the DDE contributes additional constraining power to $z\sim7$ CO model parameters.
\end{itemize}

Our numerical investigations, while preliminary, make it apparent that the non-Gaussianity of the signals imprint themselves in the DDE just as much as in the VID. Just as the the VID breaks the degeneracy inherent in the auto power spectrum between the mean CO temperature and luminosity-averaged tracer bias~\citep{Breysse22PRL}, the DDE breaks the degeneracy inherent in the cross power spectrum between the intensities of the two lines as we have demonstrated numerically. Future work should provide a non-numerical explanation for this degeneracy breaking.

The present investigation also suggests the DDE can potentially enhance science output from cross-correlations, particularly in scenarios where auto spectra and/or VID data are untrusted. Real-world application of the DDE will still need better understanding of covariance, dependence on signal and noise modelling, and so on. However, at minimum, the clearly different degeneracies from the cross power spectrum and other observables show that the DDE merits further investigation on these fronts.

\section*{Acknowledgements}

Thanks to Kieran Cleary, Tim Pearson, and other members of the COMAP collaboration that offered discussion and suggestions. Many thanks also to George Stein for running and making available the original peak-patch simulations for~\cite{Ihle19}. \added{Thanks also go to an anonymous referee for clear and actionable feedback.}

Research in Canada is supported by NSERC and CIFAR. Parts of these calculations were performed on the GPC and Niagara supercomputers at the SciNet HPC Consortium. SciNet is funded by: the Canada Foundation for Innovation under the auspices of Compute Canada; the Government of Ontario; Ontario Research Fund -- Research Excellence; and the University of Toronto.

DTC is supported by a CITA/Dunlap Institute postdoctoral fellowship. The Dunlap Institute is funded through an endowment established by the David Dunlap family and the University of Toronto. The University of Toronto operates on the traditional land of the Huron-Wendat, the Seneca, and most recently, the Mississaugas of the Credit River; DTC and others at the University of Toronto are grateful to have the opportunity to work on this land. DTC also acknowledges support through the Vincent and Beatrice Tremaine Postdoctoral Fellowship at CITA. PCB is supported by the James Arthur Postdoctoral Fellowship at New York University.

IB carried out parts of this work through the Summer Undergraduate Research Program (SURP) in astronomy and astrophysics at the University of Toronto. Work at the University of Oslo is supported by the Research Council of Norway through grants 251328 and 274990, and from the European Research Council (ERC) under the Horizon 2020 Research and Innovation Program (Grant agreement No.\ 819478, \textsc{Cosmoglobe}). Work on COMAP at Caltech is supported by the US NSF, including through NSF awards 1910999 and 2206834. The work of HP is supported by the Swiss National Science Foundation via Ambizione Grant PZ00P2\_179934. Work at Jodrell Bank is supported by an STFC Consolidated Grant (ST/T000414/1).

This research made use of Astropy,\footnote{http://www.astropy.org} a community-developed core Python package for astronomy \citep{astropy:2013, astropy:2018}. This research also made use of NASA's Astrophysics Data System Bibliographic Services.

\section*{Data Availability}

The simulated data underlying this article will be shared on reasonable request to the corresponding author.



\bibliographystyle{mnras}
\bibliography{ms,comap_es}



\appendix

\section{Correlation coefficient matrix between all observables}
\label{sec:corrcoef}

\begin{figure}
    \centering
    \includegraphics[width=0.96\linewidth,clip=True,trim=15mm 0 10mm 0]{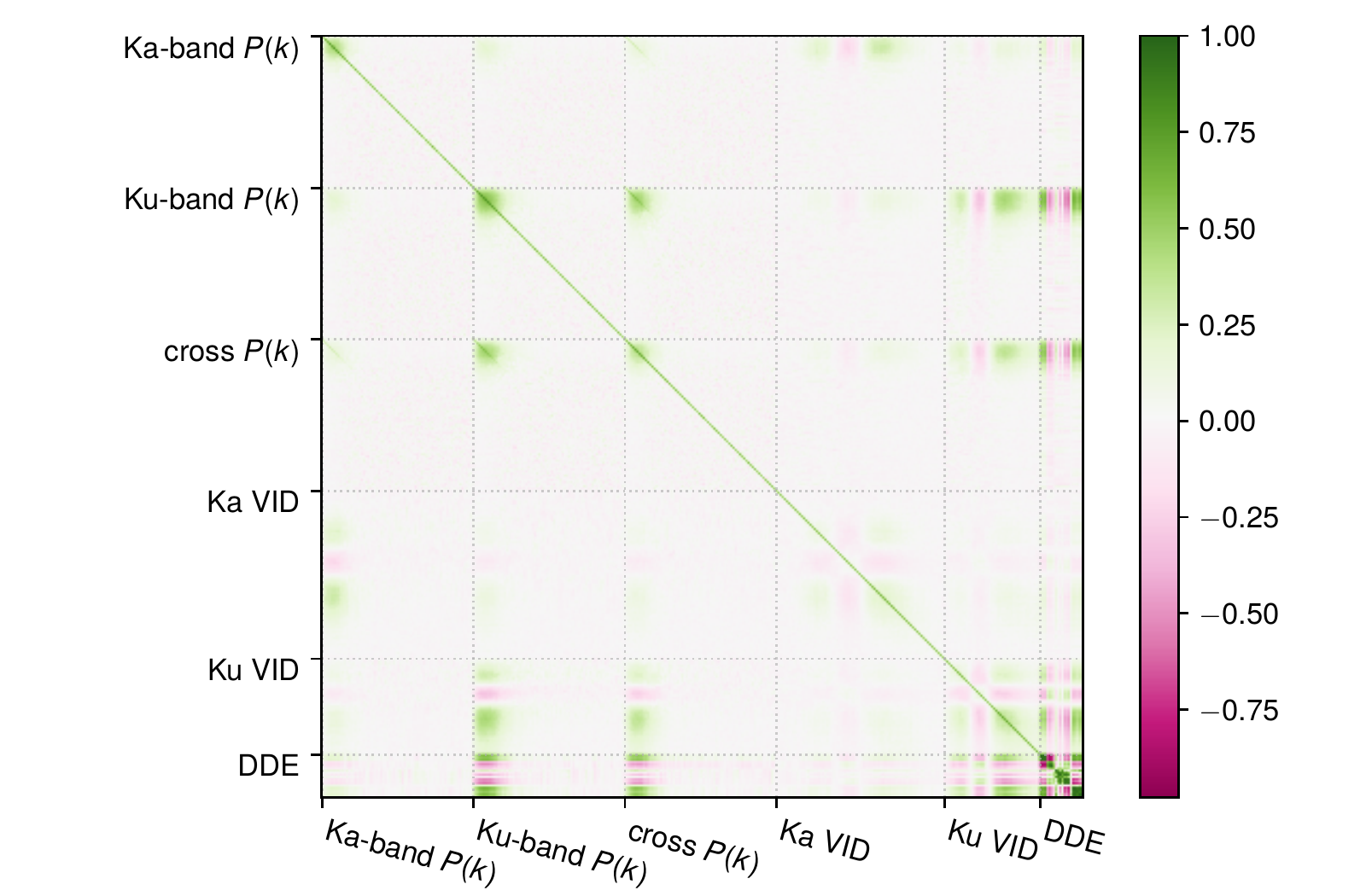}
    \caption{The correlation coefficient matrix between all observables across the 2430 simulated COMAP survey fields used in this work. We calculate and bin all observables as summarised at the end of~\autoref{sec:methods} except for the DDE, which needs to be masked as explained in~\autoref{sec:fisherres}.}
    \label{fig:corrcoef}
\end{figure}
\autoref{fig:corrcoef} shows the correlation coefficient matrix between all observables calculated in this work, across all 2430 simulations of COMAP survey fields. Aside from strong correlations between the different bins of the masked DDE (which do not affect the stability of the Fisher forecasts as explained in~\autoref{sec:fisherres}), we also find correlations between the individual VID bins and the DDE. However, correlations between the Ka-band VID and the DDE are weaker, presumably due to the interloper signal present in the Ka-band data.

Similar points apply for the low-$k$ power spectrum bins, which have correlations with VID bins likely introduced by the high-pass transfer function applied to the simulated data cubes.

\section{Tabulated Fisher forecast parameter errors}

\label{sec:fishererrors}
\begin{table*}
    \centering
    \begin{tabular}{ccccccc}
         \hline Statistics & \multicolumn{6}{c}{Marginalised Fisher forecast $1\sigma$ error for:}\\
         & $\alpha_1$ & $\beta_1$ & $\alpha_2$ & $\beta_2$ & $\alpha_{1,z\sim3}$ & $\beta_{1,z\sim3}$ \\\hline
         DDE & 0.0033 & 0.0313 & 0.0024 & 0.0225 & 0.0066 & 0.0545 \\
         $P_\times(k)$ & 0.16 & 1.46 & 0.13 & 1.19 & 0.02 & 0.47 \\
         $P_\times(k)$+DDE & 0.0030 & 0.0282 & 0.0022 & 0.0204 & 0.0056 & 0.0467\\
         VID+VID & 0.0032 & 0.0350 & 0.0095 & 0.0669 & 0.0113 & 0.1184\\
         VID+VID+DDE & 0.0018 & 0.0203 & 0.0018 & 0.0163 & 0.0041 & 0.0396 \\
         all-but-DDE & 0.0019 & 0.0215 & 0.0028 & 0.0206 & 0.0022 & 0.0224 \\
         kitchen-sink & 0.0012 & 0.0135 & 0.0013 & 0.0098 & 0.0017 & 0.0174\\\hline
    \end{tabular}
    \caption{Marginalised Fisher forecast parameter errors for each summary statistic combination considered in~\autoref{sec:fisherres}.}
    \label{tab:fishererrors}
\end{table*}

We tabulate the marginalised Fisher forecast errors for each $z\sim7$ CO--IR power law parameter in~\autoref{tab:fishererrors}, based on different sets of summary statistics. For the CO(2--1) parameters $\alpha_2$ and $\beta_2$, the DDE has constraining power equivalent to the combination of every other simulated observable combined.

\bsp	
\label{lastpage}
\end{document}